\documentclass[10pt,prd,singlecolumn,
nofootinbib, noshowkeys, noshowpacs, superscriptaddress, floatfix, aps]{revtex4-2}
\usepackage{amsmath,amsfonts,amsthm,amssymb}
\usepackage[dvips]{graphics,graphicx}
\usepackage[usenames,dvipsnames]{xcolor}
\definecolor{darkblue}{RGB}{0,0,196}
\definecolor{darkgreen}{RGB}{0,120,0}
\usepackage[colorlinks=true,linktocpage=true,linkcolor=darkblue,citecolor=red,urlcolor=darkblue]{hyperref}
\usepackage{hyperref}
\usepackage{tikz}
\usepackage{appendix}
\usepackage{pifont}
\usetikzlibrary {arrows.meta}
\usepackage[mathscr]{euscript}
\usepackage[export]{adjustbox}
\usepackage{hepunits}
\newcommand{\nn}{\nonumber}
\usepackage{stackengine,scalerel}
\newcommand\hstar[1]{\ThisStyle{\ensurestackMath{%
\setbox0=\hbox{$\SavedStyle#1$}%
\stackengine{0pt}{\copy0}{\kern.2\ht0\smash{\SavedStyle\star}}{O}{c}{F}{T}{S}}}}
\definecolor {darkgreen}{rgb}{0.2,0.7,0.2}
\begin{document}
\title{Local univalence versus stability and causality in   hydrodynamic models}
\author{Roya Heydari}
\email{rheydari@ipm.ir}
\author{Farid Taghinavaz}
\email{ftaghinavaz@ipm.ir}
\affiliation{School of Particles and Accelerator, Institute for Research in Fundamental Sciences (IPM), P. O. Box 19395-5531, Tehran, Iran.}
\begin{abstract}
Our main objective is to compare the analytic properties of hydrodynamic series with the stability and causality conditions applied to hydrodynamic modes. Analyticity, in this context, implies that the hydrodynamic series behaves as a univalent or single-valued function. Stability and causality adhere to physical constraints where hydrodynamic modes neither exhibit exponential growth nor travel faster than the speed of light. Through an examination of various hydrodynamic models, such as the Muller-Israel-Stewart (MIS) and the first-order hydro models like the BDNK (Bemfica-Disconzi-Noronha-Kovtun) model, we observe no new restrictions stemming from the analyticity limits in the shear channel of these models. However, local univalence is maintained in the sound channel of these models despite the global divergence of the hydrodynamic series. Notably, differences in the sound equations between the MIS and BDNK models lead to distinct analyticity limits. The MIS model's sound mode remains univalent at high momenta within a specific transport range.
Conversely, in the BDNK model, the univalence of the sound mode extends to intermediate momenta across all stable and causal regions. Generally, the convergence radius is independent of univalence and the given dispersion relation predominantly influences their correlation. For second-order frequency dispersions, the relationship is precise, i.e. within the convergence radius, the hydro series demonstrates univalence. However, with higher-order dispersions, the hydro series is locally univalent within certain transport regions, which may fall within or outside the stable and causal zones.
\end{abstract}
\maketitle
\section{Introduction}
Following the collisions of heavy ions in contemporary nuclear laboratories, a deconfined and many-body phase of quarks and gluons is formed, historically known as Quark-Gluon Plasma (QGP) \cite{Rischke:2003mt, Annala:2019puf, Shuryak:2004cy}. Despite its name, this state of quark matter exhibits collective behaviors that are more appropriately described using a fluid-like approach \cite{Shuryak:2003xe}. Relativistic Hydrodynamics (RH) is a fluidal description having high-energy requirements that is highly useful in explaining lead+lead collision experiments \cite{Busza:2018rrf, Heinz:2013th, Teaney:2000cw}. RH's applicability extends beyond equilibrium situations, as it is believed to be effective in far-from-equilibrium scenarios \cite{Heller:2015dha, Heller:2016rtz, Heller:2018qvh, Romatschke:2017vte}. The RH has been proposed for use in collisions of smaller particles, such as proton+proton interactions \cite{CMS:2015yux, CMS:2016fnw, ATLAS:2015hzw}. This remarkable versatility has led to the notion of the "unreasonable effectiveness of the RH" \cite{Noronha-Hostler:2015wft}, emphasizing its broad applicability and impact on the understanding of high-energy particle collisions.

In the context of high-energy systems, such as heavy-ion collisions (HIC), ensuring good performance is crucial since otherwise, the dynamic theory becomes uncontrollable, and we gain no meaningful information about the dynamics. In HIC, stability and causality serve as essential conditions that help constrain the physical space to obtain well-defined results \cite{Hiscock:1983zz}. Although the implications of these conditions on the RH series remain a debated and unsolved problem, they must still be applied. In the realm of theoretical approaches, some argue that verifying $Im \, \omega(k) \leq 0$, where $\omega$ and $k$ represent the frequency and momentum of small fluctuations, is sufficient to ensure stability. For causality, this condition is said to be met when $v = \lim\limits_{k \to \infty} \partial Re (\omega)/\partial k\leq 1$ \cite{Pu:2009fj}. Others suggest that implying $Im \, \omega(k) \leq Im (k)$ on each complex momentum is enough to have a stable and causal theory \cite{Gavassino:2023mad, Wang:2023csj, Hoult:2023clg}.

Additionally, a new trend inspired by the information current paradigm has emerged, which posits that having a non-negative divergence of the entropy current is a necessary and sufficient condition for dealing with a stable and causal theory \cite{Mullins:2023ott, Gavassino:2023odx, Gavassino:2023qwl}. This ongoing research aims to provide a better understanding of the conditions required for maintaining stability and causality in high-energy systems like heavy-ion collisions.

From a mathematical perspective, the RH series is a function of small momenta, which represents a solution of a spectral curve defined by the equation $F(\omega, k) = 0$. As a series, it raises various questions, such as convergence or divergence, obtaining finite results, exploring its analytical properties, and more. Each of these questions is significant and even more intriguing when considering their relationships with physical constraints. 

In recent years, researchers have been investigating various aspects of Relativistic Hydrodynamics (RH) to better understand its properties and applications in high-energy systems, such as heavy-ion collisions (HIC). One such area of study involves examining the radius of convergence of the RH series by solving the simultaneous equations $F(\omega, k) = 0$ and $\partial F(\omega, k)/\partial \omega = 0$ \cite{Grozdanov:2019kge, Withers:2018srf}. This approach has been successfully applied to numerous models \cite{Abbasi:2020ykq, Jansen:2020hfd, Grozdanov:2019uhi, Grozdanov:2021jfw, Asadi:2021hds, Taghinavaz:2023tog}. Another line of inquiry focuses on the analytical properties of the RH series in the context of univalent functions \cite{Grozdanov:2020koi}. Univalence is a property that pertains to an analytic function, ensuring it never takes the same value twice. Although this definition may seem simple, it imposes stringent constraints on the function. Among these constraints are the area theorem, the Bieberbach conjecture, the Littlewood limit, the growth and distortion theorem, and many others \cite{Duren:2010pm, lehto2011univalent}. 
These constraints guarantee that the analytic function is bounded and yields finite results.
By applying these limits to the transport spaces, which contain information about the underlying microscopic theory, we can better understand the constraints and properties of the RH series in high-energy systems like heavy-ion collisions \cite{Baggioli:2022uqb}. Moreover, the implications of these bounds are studied in the Quantum Field theory framework \cite{Haldar:2021rri}. This ongoing research aims to provide a deeper understanding of the mathematical foundations and properties of the RH series and its implications on the physical behavior of quark-gluon plasma and other high-energy phenomena.

Investigating the link between physical constraints like stability and causality with univalence, a mathematical limit, can provide valuable insights into the mathematical properties of the RH series and demonstrate the practical significance of these abstract limits. In this work, we explore the univalence properties of the RH series for two hydro models including the Muller-Israel-Stewart (MIS) \cite{Grozdanov:2018fic} and the first-order hydro model, known as the BDNK (Bemfica-Disconzi-Noronha-Kovtun) model \cite{Bemfica:2017wps, Kovtun:2019hdm}. Both models exhibit stability and causality in certain transport regions, making them suitable for studying the interplay between these properties and univalence. For our analysis, we focus on the Bieberbach conjecture, which bounds the size of the Taylor coefficients of the RH series and, consequently, the physical space. This conjecture can be compared with stability and causality depending on the spectral equations and the dimension of the transport space. 
 In the shear channel of the MIS and BDNK models, we observe that univalence is insignificant since the conformal map's region lies within the radius of convergence, and the original RH series remains univalent across all points within the convergence zone.

However, in the sound channel, the situation is more complex. In the MIS model, with transports like shear viscosity, relaxation time, and momentum as variables, we observe that univalence holds locally in the momentum space, even with the global divergence of the RH series. This local univalence is specified by constraint on the variable $X = -1 + \gamma_s/(8c_s^2w\tau)$, where $w = \varepsilon + p$ is the enthalpy, $c_s^2 = \partial \varepsilon/\partial p$ is the speed of sound, $\gamma_s = 4\eta/3$ is the shear viscosity, and $\tau$ is the relaxation time. The interplay between this limit and stability and causality depends on the value of $c_s$.

For the BDNK model, due to its higher-dimensional transport space and complex equations, analyzing univalence is not a straightforward process. We choose to fix the transports and vary the momentum to study where univalence occurs within stable and causal regions. Our results indicate that univalence takes place in intermediate momenta, regardless of the value of $c_s$. Nevertheless, for low and high momenta, univalence is found in the unstable and acausal zones. In conclusion, we propose that for the RH series, univalence holds locally, even though it is globally divergent. This new paradigm can be further explored in other models or for different quantities of the RH series, such as energy density, pressure, and so on.

This work is structured as follows: In Section 2, we begin by providing a comprehensive review of the fundamental concepts of univalence and its important limits. This serves as a foundation for understanding the subsequent discussions. Moving on to Section 3, we briefly introduce the background of quasihydro models, such as the Muller-Israel-Stewart (MIS) model, which is a specific example of systems with weakly broken symmetries. This section sets the stage for a more detailed analysis of these models in the following sections. In Section 4, we delve into the investigation of the relationship between stability, causality, and univalence, which is established through the Bieberbach conjecture. This section primarily focuses on the MIS model and sheds light on the interplay between these physical constraints and the mathematical limit of univalence. Section 5 extends our analysis to the BDNK model, considering its unique features and the challenges posed by its higher-dimensional transport space. In Section 6, we discuss the relationship between the radius of convergence and univalence, emphasizing the local notion of univalence.  Finally, in the concluding Section 7, we summarize the key findings of our study and outline potential avenues for future research. 
\section{Univalent functions}
A function is considered analytic when it is univalent or single-valued on a domain $D \in \mathbb{C}$, meaning it never takes the same value twice. This property defines it as a global univalent function. In contrast, a local univalent function exhibits the same behavior within the neighborhood of a point, say $\zeta_0 \in D$. By construction, univalent functions are conformal mappings due to their angle-preserving properties.
The condition $f'(\zeta_0) \neq 0$ ensures the local univalence of the point $\zeta_0$. However, a function is univalent in a convex domain iff $Re f'(\zeta) > 0$ for every point in the domain \cite{Duren:2010pm}.

Within the vast category of univalent functions, we focus on the class "$S$" due to its resemblance to the hydro series expansion. The domain of the $S$ class is the unit disk $\mathbb{D}_1 = \{\zeta : |\zeta| < 1\}$, with normalization conditions $f(\zeta=0)=0$ and $f'(\zeta=0) = 1$. There exists a general argument that for a given univalent function $f(z)$ defined on a simply connected domain $U \in \mathbb{C}$, a reversible map such as $\zeta = \varphi(z)$ can be defined that entails smooth transformation of any $z \in U$ to $\zeta \in \mathbb{D}_1$ and consequently $f(z) \to g(\zeta) = f(\varphi^{-1}(\zeta))$  in which $g(\zeta) \in S$. This transformation is achieved using the Riemann mapping theorem. A general series expansion for the $S$ class univalent function $f(\zeta)$, can be represented as follows:
\begin{align}\label{eqs}
    f(\zeta) = \zeta + \sum_{n=2}^{\infty} a_n \, \zeta^n, \qquad \vert \zeta \vert \leq 1.
\end{align}

The univalence of a function imposes stringent constraints on the size of its Taylor coefficients. These constraints originate from the well-known area theorem \cite{Duren:2010pm}
\begin{align}\label{eq: area}
    \sum_{n=1}^\infty \, n \vert b_n \vert ^2 \leq 1,
\end{align}
where $b_n$ are the coefficients for a function $h(\zeta)$ belonging to the class $\Sigma$ of univalent functions, which can be represented as follows
\begin{align}
    h(\zeta) = \zeta + b_0 + \sum_{n=1}^\infty \, b_n \, \zeta^{-n}.
\end{align}
The class $\Sigma$ functions are defined on the exterior domain of the unit disk $\mathbb{D}_1$ and map a region, say $\Delta$, to the complement of a compact, connected set. The relationship between the class $\Sigma$ and $S$ can be established through a suitable inverse transformation from where the following constraints can be obtained.

Inspired by the area theorem provided in Eq. \eqref{eq: area}, Bieberbach conjectured that the expansion coefficients of univalent $S$ functions in Eq. \eqref{eqs} are bounded
\begin{align}\label{eqb}
    \vert a_n \vert \leq n, \,\,\, n \geq 2.
\end{align}
This conjecture has been proven for $n=2, 3, 4, 5$, and $6$, and for all $n$ within certain subclasses of $S$. However, the full conjecture remains an open problem. Numerous constraints exist, some of which have been proven concretely. For instance, Littlewood's theorem states that for each $f \in S$, we have \cite{Duren:2010pm}
\begin{align}\label{eq: little}
   \vert a_n \vert \leq e \, n, \qquad n \geq 2.
\end{align}
The area theorem has far-reaching consequences, including the "growth theorem" for each function $f \in S$ \cite{Duren:2010pm}
\begin{align}\label{eqg}
    \frac{\vert \zeta \vert}{(1 + \vert \zeta \vert)^2} \leq \vert f(\zeta)\vert \leq \frac{\vert \zeta \vert }{(1 - \vert \zeta \vert)^2}.
\end{align}
This theorem demonstrates that the absolute value of any univalent $S$ function on each point of the unit disk is bounded, implying that the series in Eq. \eqref{eqs} converges for all points within the unit disk. The area theorem also yields other results, such as the "distortion theorem," which can be derived from the derivatives of Eq. \eqref{eqg}. It is worth mentioning that both conditions \eqref{eqb} and \eqref{eqg} are saturated by the so-called Koebe function
\begin{align}
    f(\zeta) = \frac{\zeta}{(1 - \zeta)^2} = \sum_{n=1}^\infty n \zeta^n.
\end{align}
This function conformally maps $\mathbb{D}_1$ to $\mathbb{C} \backslash (-\infty, -1/4)$ and is of great importance in univalent studies. 

Having provided the necessary background, we now aim to apply the aforementioned process to relativistic hydro models. To achieve this, we must follow several steps:
\begin{enumerate}
\item  Analyze the analytical properties of the hydro series in the complex $z \equiv k^2$ plane. We need to determine whether the univalence condition is satisfied by ensuring Re$f'(z) > 0$.

\item  Construct a conformal map $\zeta = \varphi(z)$ that transforms the region $U$ to $\mathbb{D}_1$. This step is crucial, as it requires a thorough understanding of the complex $z$ analytical structure, including poles, branch points, branch cuts, and other relevant features.

\item  Apply the map derived in step 2 to the original hydro series to obtain the $a_n$ coefficients in Eq. \eqref{eqs} and examine the constraints given by Eqs. \eqref{eqb} or \eqref{eq: little}.
\end{enumerate}

This procedure has been studied for normal hydro expansions in \cite{Grozdanov:2020koi} and for a specific holographic model in \cite{Baggioli:2022uqb}. Our goal is to apply these methods to particular models, such as the MIS or BDNK models. Both of these models share a common characteristic: they are causal and stable in certain regions of transport space. Investigating the interplay between analyticity constraints, stability, and causality can reveal both the mathematical and physical properties of the hydro series. Moreover, this comparison will help us understand the significance of the univalence limits.
\section{Review of hydrodynamic models}
\subsection{Quasihydrodynamic models}
Quasihydro models, or "systems with weakly broken symmetries," refer to systems where the conserved current dynamics alone are insufficient to describe the entire system's dynamics. Hydrodynamics is a theory applicable to length and time scales that are long compared to the mean free time $\tau_{mft}$ and mean free path $\ell_{mfp}$ \cite{Grozdanov:2018fic}. So, the gradient expansion is valid for scales beyond these
\begin{align}
    \tau_{mft} \partial_t \ll 1, \qquad \ell_{mfp} \partial_x \ll 1.
\end{align}
In these regimes, conserved currents are responsible for describing long-range thermal excitations
\begin{align}
     \partial_t \langle \rho \rangle + \partial_i J^i = 0.
\end{align}
Now, suppose there exists a non-conserved operator $\mathcal{P}$ in the theory's spectrum, with a lifetime of order $\tau_1$. This can occur for other non-conserved operators $\mathcal{P}_i$ with lifetimes $\tau_i$. We can assume that the time scales of non-conserved operators follow a hierarchy such as $\tau_1 \gg (\tau_2, \tau_3, \cdots)$ and $\tau_1 \lesssim \tau_{mft}$. In such a setting, the hydro series loses its validity when $\tau_1 \partial_t \sim 1$, as on these time scales, other modes contribute, including low-lying modes of $\mathcal{P}$ or the first non-hydro modes. The typical equations for this setup are
\begin{align}\label{eqt}
     &\partial_t \langle \rho \rangle + \partial_i J^i = 0,\nn\\
     &\partial_t \langle \mathcal{P} \rangle + \partial_i J^i_P = -\frac{\langle \mathcal{P} \rangle}{\tau_1}.
\end{align}
The first equation represents the conservation of energy and momentum charges, while the second demonstrates how non-hydro modes relax to equilibrium values through a relaxation equation. If $\tau_2$ were the next smallest relaxation time, Eq. \eqref{eqt} would be valid up to $\tau_2 \partial_t < 1$. In time scales of order $\tau_2 \partial_t \simeq 1$, the next non-hydro mode comes into play, and so on. Indeed, non-conserved operators serve as regulator fields to compensate for the incompleteness of conserved operators.

 The simplest quasihydro model is the diffusion-to-sound model, which interpolates from Fick's law at small momentum to a propagating sound mode at large momentum \cite{Grozdanov:2018fic}. Traditional Fick's law for the diffusion of a locally conserved current $S$ can be written as
\begin{align}
     \partial_t S = D \, \partial_x^2 S.
\end{align}
The dispersion relation of this model is given by $\omega = - i D\, k^2$, which violates stability and causality. To overcome this issue, we introduce an auxiliary field $J$, which plays a regulatory role
\begin{align}
     &\partial_t S + \partial_x J =0, \nn\\
     &\partial_t J + \frac{D}{\tau} \partial_x S = - \frac{J}{\tau}.
\end{align}
The dispersion relation of the modes is given by Eq. \eqref{eq10}
\begin{align}\label{eq10}
     F(\omega, k^2) &=  \,\,\omega^2 + \frac{i \, \omega}{\tau} - \frac{D \, k^2}{\tau} = 0,
\end{align}
which has the solutions
\begin{align}
     \omega_\pm &= -\frac{i}{2 \tau} \left(1 \pm \sqrt{1 - 4 D \tau k^2}\right).
\end{align}
It is easy to show that the small and large $k$ limits of the modes are obtained as
\begin{align}
     \mbox{Small $k$}, & \quad \omega_+ = -\frac{i}{\tau}+ i D k^2 + \cdots, \qquad \omega_- = - i D k^2 + \cdots,\nn\\
     \mbox{Large $k$}, & \quad \omega_\pm = \pm \sqrt{\frac{D}{\tau}} k - \frac{i}{2 \tau} + \mathcal{O}(1/k).
\end{align}
At small $k$, we get two damping modes, while the large $k$ features the propagating behavior. This demonstrates that at a certain point, $k_c$, the diffusive modes turn into propagating ones. This change occurs at $k_c^2= 1/(4\tau D)$. The causality criteria at large $k$ are given by Eq. \eqref{eq12} \cite{Pu:2009fj}
\begin{align}\label{eq12}
     v = \lim_{k \to \infty} \bigg\vert \frac{\partial Re (\omega)}{\partial k}\bigg\vert = \sqrt{\frac{D}{\tau}} \leq 1, \,\,\, \mbox{leads to} \quad 0 \leq D \leq \tau.
\end{align}
Using the Routh-Hurwitz criteria for stability in any $k$ ensures non-negative transports. This limit is also obtained by low $k$ limits. It is noteworthy to mention that the point $k_c$ represents the location of mode collision, as $F(\omega_\pm, k^2) = 0$ and $\frac{\partial F(\omega, k^2) }{\partial \omega}\bigg|_{\omega = \omega_\pm} = 0$ \cite{Grozdanov:2018fic, Grozdanov:2019kge}. Many examples lie within the quasihydro model context, such as the Muller-Israel-Stewart (MIS), Magnetohydrodynamics, or systems with broken symmetries, which can be studied similarly to the diffusion-to-sound model.
\subsection{BDNK model}
The essence of the BDNK or first-order hydro model is to not use the regulator fields \cite{Kovtun:2019hdm, Bemfica:2017wps}. Instead, we restrict ourselves to the hydro fields $(u^\mu, T, \mu)$, but a modification is made in the constitutive relations. This process works as follows. Generally, the energy-momentum tensor and charge currents of an out-off-equilibrium fluid are written as
\begin{align}\label{eqtj}
    &T^{\mu \nu} = \mathcal{E} u^\mu u^\nu -\mathcal{P} \Delta^{\mu \nu} + \mathcal{Q}^\mu u^\nu + \mathcal{Q}^\nu u^\mu + \mathcal{T}^{\mu \nu},\nn\\
    &J^\mu = \mathcal{N} u^\mu + \mathcal{J}^\mu,
\end{align}
where $\Delta^{\mu \nu} = \eta^{\mu \nu} - u^\mu u^\nu$ is the projection tensor onto the perpendicular surface of $u^\mu$ and $u^\mu u_\mu = 1$. The $\mathcal{Q}^\mu$ and $\mathcal{J}^\mu$ vectors as well as the $\mathcal{T}^{\mu \nu}$ tensor are transverse to $u^\mu$. The dynamical equations are conservation laws of energy, momentum, and $U(1)$ charge density
\begin{align}\label{eqce}
    &\partial_\mu T^{\mu \nu} =0,\nn\\
    &\partial_\mu J^\mu = 0.
\end{align}
At zeroth order of derivatives, $\mathcal{E}, \mathcal{P}, \mathcal{N}, \mathcal{Q}^\mu, \mathcal{J}^\mu$ and $\mathcal{T}^{\mu \nu}$ equal to their equilibrium values. In higher orders of derivatives, they get some corrections due to the out-off-equilibrium fluctuations. The BDNK approach gives these corrections to the hydro fields profiles or their derivatives up to a desired level by considering the symmetries of functions \cite{Kovtun:2019hdm}
\begin{align}\label{eqcr}
  & \mathcal{E} = \varepsilon + \sum_{n=1}^\infty \sum_{i=1}^k \epsilon_i s_n^i, \quad \mathcal{P} = p + \sum_{n=1}^\infty \sum_{i=1}^k \pi_i s_n^i, \quad \mathcal{N} = n + \sum_{n=1}^\infty \sum_{i=1}^k \nu_i s_n^i,\nn\\  
  & \mathcal{Q}^\mu = \sum_{n=1}^\infty \sum_{i=1}^k \bar{\theta}_i v_n^{\mu   i}, \quad \mathcal{J}^\mu = \sum_{n=1}^\infty \sum_{i=1}^k \gamma_i v_n^{\mu  i}, \nn\\
  &\mathcal{T}^{\mu \nu} = \sum_{n=1}^\infty \sum_{i=1}^k \eta_i t_n^{\mu \nu  i}.
\end{align}
The index $n$ counts the derivative order and the index $i$ runs through the available and independent sets of scalar $s^i_n$, vector $v_n^{\mu i}$, and tensor $t_n^{\mu \nu i}$ bases. Lists of these bases can be found in \cite{Jensen:2012jh}. The new transport coefficients $(\epsilon_i, \pi_i, \nu_i, \bar{\theta}_i, \gamma_i, \eta_i)$ quantify the amounts of getting away from zeroth order. The interesting feature of this model is that the redefinition of hydro fields
\begin{align}
    T \to T + \delta T, \quad u^\mu \to u^\mu + \delta u^\mu, \quad \mu \to \mu + \delta \mu,
\end{align}
can be recast as changes of the transports \cite{Kovtun:2019hdm}. Likewise, to implement this model, it is crucial to impose the constraints results from having a well-defined thermodynamic state as follows \cite{Jensen:2012jh}
\begin{align}\label{eq: thermo}
    & \partial_\mu T + T a_\mu = 0, \nn\\
    &\partial_\mu \mu + \mu a_\mu = E_\mu, \nn\\
    & \nabla_\mu u_\nu + u_\mu a_\nu = \omega_{\mu \nu},   
\end{align}
where
\begin{align}
    a_\mu = u^\nu \nabla_\nu u_\mu, \qquad \nabla_\mu = \Delta_{\mu \alpha} \partial^\alpha, \qquad \omega_{\mu \nu} = \frac{\nabla_\mu u_\nu - \nabla_\nu u_\mu}{2},
\end{align}
and $E_\mu = F_{\mu \nu} u^\nu$ is the electric field in the reference frame of the fluid, $a_\mu$ is the acceleration vector, $\nabla_\mu$ is the perpendicular derivative and $\omega_{\mu \nu}$ is the vorticity. Having well-defined thermodynamics that is parameterized in Eq. \eqref{eq: thermo} will reduce the number of transport coefficients and keep only the independent ones \cite{Jensen:2012jh}. By considering the relations \eqref{eqtj} and \eqref{eqcr} and plugging into the Eq. \eqref{eqce}, we establish a mathematical basis to analyze the stability and causality of hydro modes for a particular system with a definite equation of state. The outcomes of these analyses are essentially limits on the parameter space of independent transports \cite{Kovtun:2019hdm, Hoult:2020eho, Taghinavaz:2020axp}.

Having reviewed the necessary models, we proceed to analyze the interplay between constraints derived from stability and causality with the univalence bounds in the MIS and the BDNK model. We aim to explore whether these constraints overlap, shedding light on the significance of univalence bounds in the generation of new physical constraints. We will discover that univalence holds locally, meaning that in specific regions of momentum space, bounded results can be achieved. This observation is observed for series that are not globally convergent.
 \section{Interplay of bounds in the MIS model}
 In the context of the MIS model, we refer to an uncharged conformal system with the following constitutive relation
 \begin{align}
     T^{\mu \nu} = \varepsilon u^\mu u^\nu - p \Delta^{\mu \nu} + \Pi^{\mu \nu},
 \end{align}
 where $\Pi^{\mu \nu} = -2 \eta \sigma^{\mu \nu}$ with $\eta$ denoting the shear viscosity and $\sigma^{\mu \nu} = \frac{1}{2} \left(\nabla^\mu u^\nu + \nabla^\nu u^\mu - \frac{2}{3} \Delta^{\mu \nu} \nabla \cdot u\right)$ represents the shear-stress tensor. The governing equations are given by
 \begin{align}\label{eq: eqs-MIS}
     &\partial_\mu T^{\mu \nu} = 0,\nn\\
     & \tau u^\nu \partial_\nu \Pi^{\mu \nu} + \Pi^{\mu \nu} = - 2 \eta \sigma^{\mu \nu},
 \end{align}
 where $\tau$ is the relaxation time associated with $\Pi^{\mu \nu}$ approaching its on-shell value, $-2\eta\sigma^{\mu \nu}$. Eqs. \eqref{eq: eqs-MIS} resemble the Eq. \eqref{eqt} and $\Pi^{\mu \nu}$ serves as a regulator field with the lifetime $\tau$, being the first non-conserved operator in the hydro's spectrum.  Solving the Eqs. \eqref{eq: eqs-MIS} for small perturbations and decomposing them into the shear and sound channel the following dispersion equations are derived
 \begin{align}
     \mbox{Shear:} \qquad & \omega^2 + i \frac{\omega}{\tau} - \frac{\eta k_z^2}{w \tau} = 0,\\
     \mbox{Sound:} \qquad & \omega^3 + \frac{i \omega^2}{\tau} - \left(c_s^2 + \frac{\gamma_s}{w \tau}\right) \omega k_z^2 - \frac{i c_s^2 k_z^2}{\tau}= 0,
 \end{align}
 where $c_s^2 = \partial \varepsilon/\partial p$ denotes the speed of sound, $\gamma_s = 4\eta/3$ and $w = \varepsilon + p$ is enthalpy. In this section, we will meticulously analyze the aforementioned equations to examine their analytical properties and compare them with the prerequisites of stability and causality. By understanding the connections between these equations and the fundamental physical constraints, we can gain valuable insights into the behavior and limitations of the MIS model.
 \subsection{Shear channel}
In the shear channel, the dispersion relation for fluctuations in the transverse plane can be represented as
 \begin{align}\label{eq13}
     F^{\text{MIS}}_\text{shear}(\omega, z ) = \omega^2 + i \frac{\omega}{\tau} - \frac{\eta z}{w \tau} = 0,
 \end{align}
 where $z = k_z^2$. To derive the stability constraints, we apply the Routh-Hurwitz criterion, assuming $\omega(z) = i \beta(z)$. The resulting coefficients must satisfy the Routh-Hurwitz conditions, which leads to the following dynamical stability conditions
 \begin{align}
     \tau >0, \quad \eta >0.
 \end{align}
 Additionally, we require $w > 0$ as a thermodynamic stability condition.  The causality constraint is given by the large momenta behaviors mentioned in Eq. \eqref{eq12}, and results in  $\eta \leq w \, \tau$. These constraints ensure the model's behavior adheres to the principles of stability and causality.

\begin{figure}
    \centering
    \includegraphics[width=0.6\textwidth]{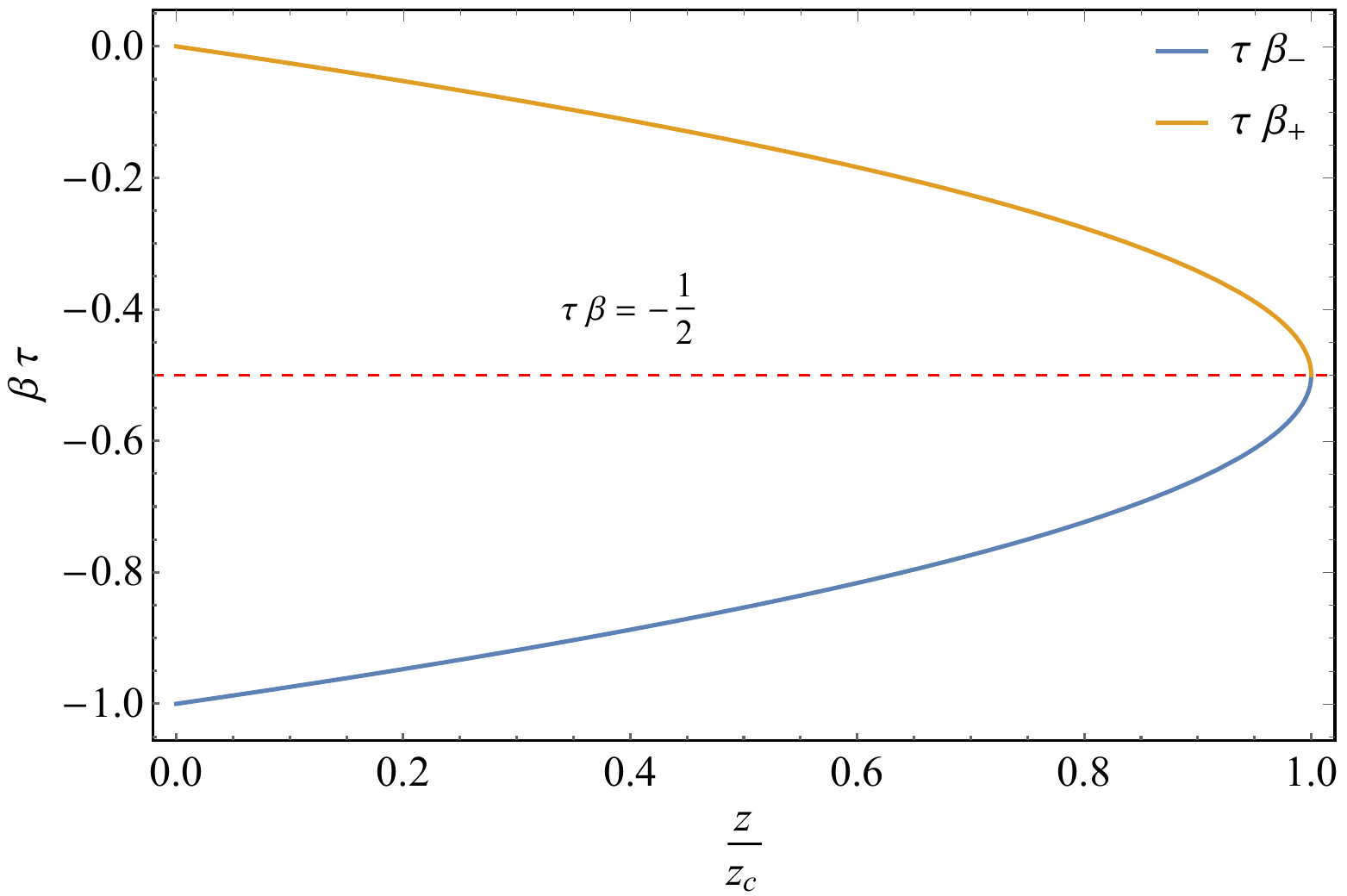}
    \caption{Plot of modes in terms of $\frac{z}{z_c}$ given in Eq. \eqref{eq16}. The red dashed line indicates the upper limit for which the univalence condition $\mbox{Re} (\beta'(z)) \geq 0$ holds.}
    \label{fig1}
\end{figure}

To derive the univalence bound, we must first verify the condition $\text{Re}(\beta'(z)) \geq 0$, which is necessary and sufficient for a function to be univalent. Applying this condition to Eq. \eqref{eq13},  we obtain the following expression for the derivative 
 \begin{align}\label{eq: betap-MIS-sol}
     \beta'(z) = - \frac{\eta}{w \left(1 + 2 \tau \beta(z)\right)}.
 \end{align}
 For the function to be univalent, we need solutions with $\beta(z) < -\frac{1}{2\tau}$. In Eq. \eqref{eq: betap-MIS-sol}, the $\beta(z)$ are solutions of  Eq. \eqref{eq13}
 \begin{align}\label{eq16}
     \beta_{\pm}(z) = \frac{1}{2 \tau} \left( - 1 \pm \sqrt{1 - \frac{4 z \eta \tau}{w}} \right).     
 \end{align}
 In Fig. \ref{fig1}, we plot these solutions in terms of dimensionless quantities, specifically $\beta \tau$ in terms of $z/z_c$. From this figure, we infer that the blue branch or $\tau \beta_-$ always satisfies $\beta(z) < - \frac{1}{2 \tau}$. Thus we must take this solution as a univalent function. The red dashed line represents $\tau \beta = - 1/2$, and the values below it are acceptable due to the univalence condition. The point $z_c = \frac{w}{4 \eta \tau}$ marks where the modes intersect, setting a maximum bound for applying the hydro series to the $\beta(z)$ solutions.
To analyze the shear channel further, we employ a reversible and conformal map $\zeta = \phi(z)$ to transform the $z$ space into the unit disk $\mathbb{D}_1$. The solutions in Eq. \eqref{eq16} exhibit a branch cut along $[z_c, \infty)$. According to the prescription given in \cite{Grozdanov:2020koi}, the map is provided as
\begin{align}\label{eq: map-shear}
    z = \phi_{\text{shear}}^{-1}(\zeta) = - \frac{4 z_c \zeta}{(1- \zeta)^2}.
\end{align}
By substituting this map into the $\tau\beta_-$ from Eq. \eqref{eq16}, we obtain
\begin{align}
    f_{\text{shear}}(\zeta) = \tau \, \beta(\phi_{\text{shear}}^{-1}(\zeta)) = - \frac{1}{1 - \zeta}.
\end{align}
This function is unrelated to the transports and remains univalent for all points within the unit disk $\mathbb{D}_1$. Consequently, the Bieberbach conjecture cannot impose constraints on the model's transports in the shear channel. This observation leads us to conclude that, in this channel, all constraints stem from stability and causality, while univalence does not introduce any additional conditions. This phenomenon likely occurs because the radius of convergence and the map's domain are equivalent in the shear channel. For further details, refer to Section \ref{sec: rel-SC-U}.

\subsection{Sound channel}
As mentioned earlier, the dispersion relation for the sound channel is given as \cite{Grozdanov:2018fic}
\begin{align}\label{eq: disp-sc-QH}
    F^{\text{MIS}}_\text{sound}(\omega, z = k_z^2) = \omega^3 + \frac{i \omega^2}{\tau} - \left(c_s^2 + \frac{\gamma_s}{w \tau}\right) \omega z - \frac{i c_s^2 z}{\tau}= 0,
\end{align}
The Routh-Hurwitz limits ensure the stability of perturbations at each $z$. It demands $Re \,(\omega)>0$ for each $z$ and results to
\begin{align}\label{eq: sc-stab-QH}
    \tau > 0, \quad \eta > 0.
\end{align}
The causality criterion is obtained by examining the large $k_z$ expansion of the dispersion relation Eq.  \eqref{eq: disp-sc-QH}.  Plugging $\omega = v k_z$ into Eq. \eqref{eq: disp-sc-QH} and setting the coefficient of the dominant power of $k$ to zero, we arrive at
\begin{align}\label{eqcausal}
    &v = \lim_{k \to \infty} \bigg|\frac{\partial Re (\omega)}{\partial k}\bigg| = \sqrt{c_s^2 + \frac{\gamma_s}{w \tau}} \leq 1,\quad \Rightarrow \,\,\, \frac{\gamma_s}{w \tau} \leq 1 - c_s^2.
\end{align}

\begin{figure}
    \centering
    \includegraphics[width=0.7\textwidth]{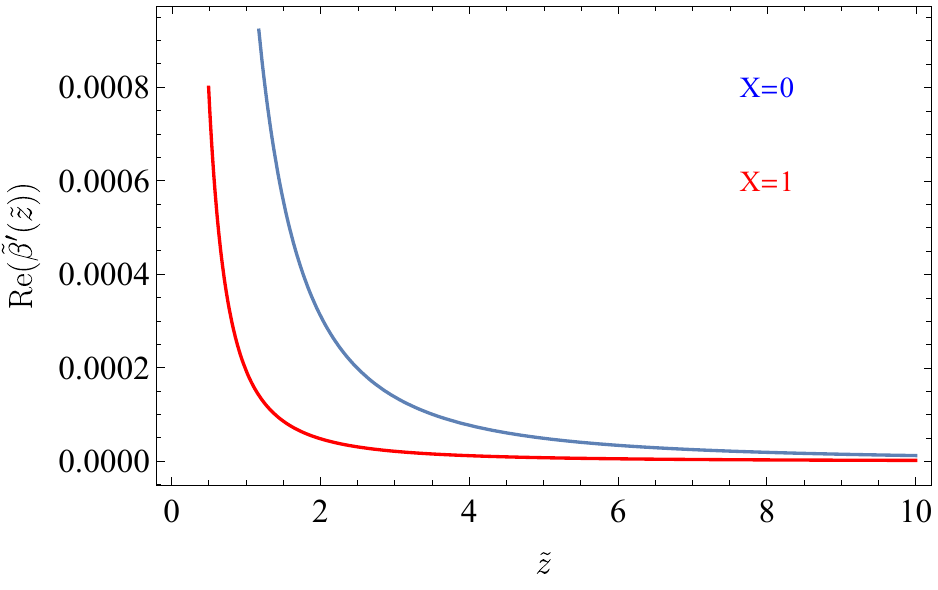}
    \caption{Plot of $Re \tilde{\beta}'(\tilde{z})$ from Eq.  \eqref{eq: betap-sc-QH} for $\tilde{\beta_1}(\tilde{z})$ solution provided in Eq. \eqref{eq: sound-sol-QH}. The red (blue) plot refers to the choice $X = 1(X=0)$.}
    \label{fig2}
\end{figure}

To investigate the univalence bounds, we first need to check the validity of $Re \, (\beta'(z)) > 0$ which can be obtained from Eq.  \eqref{eq: disp-sc-QH} as follows
\begin{align}\label{eq: betap-sc-QH}
    \tilde{\beta}'(\tilde{z}) = \frac{\partial \tilde{\beta}}{\partial \tilde{z}} = - \frac{1 + (9 + 8 X) \tilde{\beta}(\tilde{z})}{(9+ 8 X)\tilde{z} +  \tilde{\beta}(\tilde{z}) (2 + 3 \tilde{\beta}(\tilde{z}))} \geq 0,
\end{align}
where
\begin{align}\label{eq: dimless-sc-QH}
    X \equiv -1 + \frac{\gamma_s}{8 c_s^2 w \tau}, \qquad \tilde{z} \equiv c_s^2 \tau^2 z, \qquad \tilde{\beta} \equiv \tau \, \beta .
\end{align}
To proceed, we must evaluate the Eq.  \eqref{eq: betap-sc-QH} on the solutions of Eq. \eqref{eq: disp-sc-QH}. These solutions are given by
\begin{align}\label{eq: sound-sol-QH}
    &\tilde{\beta_1}(\tilde{z}) = - \frac{1}{3} \bigg(1 + \frac{3 \tilde{z} (9 + 8 X) - 1}{\left(\sqrt{\mathcal{Q}_1(\tilde{z})} + 9 \tilde{z} (3 + 4 X) - 1\right)^{1/3}} - \left(\sqrt{\mathcal{Q}_1(\tilde{z}) } + 9 \tilde{z} (3 + 4 X) - 1\right)^{1/3}\bigg),\\
    &\tilde{\beta}_{s}(\tilde{z}) = - \frac{1}{6} \bigg(2 - \frac{(1 + s i \sqrt{3})\left(3 \tilde{z} (9 + 8 X) - 1\right)}{\left(\sqrt{\mathcal{Q}_1(\tilde{z})} + 9 \tilde{z} (3 + 4 X) - 1\right)^{1/3}} + (1 - s i \sqrt{3})\left(\sqrt{\mathcal{Q}_1(\tilde{z}) } + 9 \tilde{z} (3 + 4 X) - 1\right)^{1/3}\bigg),\nn
\end{align}
where
\begin{align}
    \mathcal{Q}_1(\tilde{z}) \equiv \left(3 \tilde{z} (9 + 8 X) - 1\right)^3 + \left(9 \tilde{z} (3 + 4 X) - 1\right)^2.
\end{align}
and $s = \pm $. Out of the solutions in Eq.  \eqref{eq: sound-sol-QH}, only $\tilde{\beta}_1(\tilde{z})$ satisfies the condition $Re (\tilde{\beta}'(\tilde{z})) > 0$. The others result in negative values or yield no result in the physical region. In Fig. \ref{fig2}, we plot the $Re (\tilde{\beta}'(\tilde{z}))$ by substituting $\tilde{\beta}_1(\tilde{z})$ from Eq.  \eqref{eq: sound-sol-QH} into it for $X=0$ and $X=1$. For other choices of $X$, we always have $Re (\tilde{\beta}'(\tilde{z})) > 0$. The significant advantage of the dimensionless equation using the notations in \eqref{eq: dimless-sc-QH} is that there is only one parameter $X = -1 + \frac{\gamma_s}{8 c_s^2 w \tau}$ to label the solutions. It is worth mentioning that stability conditions in Eq.  \eqref{eq: sc-stab-QH} require $X \geq -1$.

\begin{figure}
\centering
\begin{tikzpicture}[scale=0.9]
\draw[->] (-4,0) -- (4,0) node[below, yshift = -0.10cm, scale = 1.25]{\textbf{Re $\tilde{z}$}};
\draw [->](0,-4) -- (0,4) node[above,  xshift = -0.65cm, yshift = -0.45cm, scale = 1.25]{\textbf{Im $\tilde{z}$}};
\draw [black, thick](1,-4) -- (1,4);
\draw  [red, thick, arrows = {-Stealth[]}](1.25,1) -- (3.75,1) 
node[above, yshift = 0.15cm, scale = 1.2]{HM};
\draw  [black, dashed, ultra thick] (2,0) -- (4,0);
\draw [blue, thick, arrows = {-Stealth[]}](0.75 ,1) -- (-2,1) 
node[above, yshift = 0.15cm, scale = 1.2]{LM};
\draw  [black, dashed, ultra thick] (0,0) -- (1,0);
\filldraw [black] (0,0) circle (3pt) node[below, xshift = -0.2cm, yshift = -0.05cm, scale = 1.25]{$\tilde{z}_0$};
\filldraw [black] (1,0) circle (3pt) 
node[below, xshift = -0.25cm, yshift = -0.05cm, scale = 1.25]{$\tilde{z}_-$};
\filldraw [black] (2,0) circle (3pt) node[below , yshift = -0.05cm, scale = 1.25]{$\tilde{z}_+$};
\end{tikzpicture}
\caption{Singularity pattern of solutions in complex $z$ plane for Eq.  \eqref{eq: disp-sc-QH}. The HM(High Momentum) and LM (Low Momentum) regions are separated by a vertical solid line on $\tilde{z}_-$. The dashed lines on the horizontal Re $\tilde{z}$ line refer to the branch cuts. The black dots mention the symbolic locations of $\tilde{z}_0$ and $\tilde{z}_\pm$ given in Eq. \eqref{eq: tilz-sol-QH}.}\label{fig: sing-sc-QH}
\end{figure}
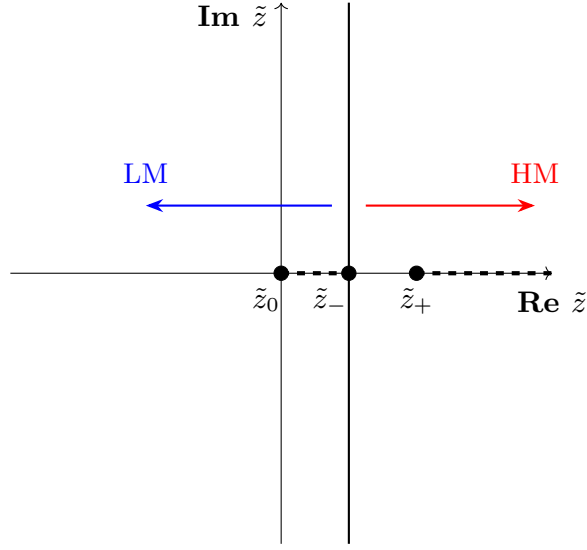

To derive the conformal map relating $z$ to the $\xi$ plane, we need to understand the analytical structure of the solutions. The singularity pattern is a collection of branch cuts due to the square root appearance of $\mathcal{Q}_1(z)$.
\footnote{Indeed, the branch cuts are determined by line segments where below and above them, there is a $"\pi"$ phase difference.} The branch points are roots of $\mathcal{Q}_1(z) = 0$ which are given as follows
\begin{align}\label{eq: tilz-sol-QH}
    \tilde{z}_0 = 0, \qquad \tilde{z}_\pm = \frac{9 (3 + 4 X) + 8 X^2 \pm 8 \sqrt{X^3  (1+X)}}{\left(9 + 8 X \right)^3}.
\end{align}
At $\tilde{z} = \tilde{z}_0$ we get $\tilde{\beta}_1(\tilde{z}_0) = \tilde{\beta}_-(\tilde{z}_0) = 0$, at $\tilde{z} = \tilde{z}_-$ we get $\tilde{\beta}_1(\tilde{z}_-) = \tilde{\beta}_-(\tilde{z}_-)$, and at $\tilde{z} = \tilde{z}_+$ we get $\tilde{\beta}_-(\tilde{z}_+) = \tilde{\beta}_+(\tilde{z}_+)$. 
In Fig. \ref{fig: sing-sc-QH}, the analytical structure of solutions in the Eq.  \eqref{eq: sound-sol-QH} is sketched in the complex $\tilde{z}$ plane. There are two branch cuts one in $0 \leq \tilde{z} \leq \tilde{z}_-$ and the other in $\tilde{z} \geq \tilde{z}_+$.  Due to these disjoint cuts, it is impossible to introduce a unique conformal map from $\tilde{z}$ to the $\xi$ plane. So we break the $\tilde{z}$ plane into the HM (High Momentum) $\tilde{z} \geq \tilde{z}_-$, and LM (Low Momentum) $\tilde{z} \leq \tilde{z}_-$ regions which are separated by a vertical solid line. In the HM region, the following map is chosen  
\begin{align}\label{eq: map-HM-QH}
    -\frac{\tilde{z}}{4(\tilde{z}-\tilde{z}_+)} = \frac{\zeta}{(1-\zeta)^2}, \,\,\, \Rightarrow \quad \tilde{z} = \phi_{(HM)}^{-1}(\zeta) = \frac{4 \tilde{z}_+ \zeta}{(1+\zeta)^2}.
\end{align}
This map is obtained by first sending the branch-cut $\tilde{z}\geq \tilde{z}_+$ into the $(-\infty, -\frac{1}{4}]$ and then using it as the range of the Koebe function. In the LM region, we employ the following map
\begin{align}\label{eq: map-LM-QH}
    -\frac{\tilde{z}_-}{4 \tilde{z}} = \frac{\zeta}{(1-\zeta)^2}, \,\, \Rightarrow \quad \tilde{z} = \phi_{(LM)}^{-1}(\zeta) = -\frac{\tilde{z}_- (1-\zeta)^2}{4 \zeta},
\end{align}
 where the branch-cut $0\leq \tilde{z} \leq \tilde{z}_-$ sends to the $(-\infty, -\frac{1}{4}]$ and then uses it as the range of the Koebe function.  Each map has to be inserted in $\tilde{\beta}_1(\tilde{z})$ and expanded to check the univalence condition. 

\begin{figure}
    \centering
    \includegraphics[width=0.5\textwidth]{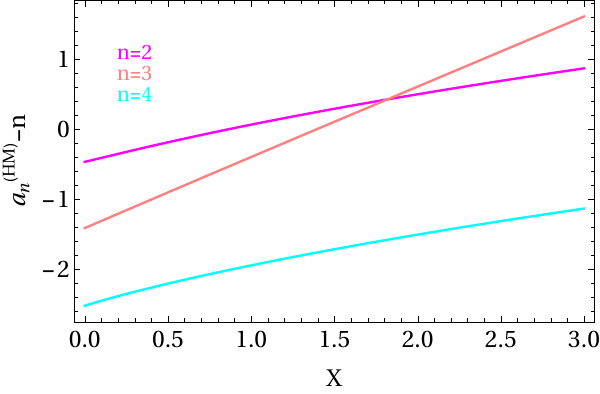}
    \caption{Plot of the Bieberbach conjectures, $a_n^{(HM)} -n$ for various $n$ in terms of $X = -1 + \gamma_s/(8c_s^2 w \tau)$ in the HM region of the analytical plane. For $X \leq 0.853$ the Bieberbach conjecture, i.e. $a_n \leq n$ is satisfied for all curves.}
    \label{fig: a2-2-p1-sound}
\end{figure}

In the HM Region, the map \eqref{eq: map-HM-QH} is inserted into the $\tilde{\beta}_1(\tilde{z})$ and after a bit of calculation we get the following series 
\begin{align}\label{eq: soundunivalence-1}
    &f^{HM}_{\text{sound}}(\zeta) =  \tilde{\beta}_1(\phi_{(HM)}^{-1}(\zeta)) = \zeta + a_2^{(HM)} \zeta^2 + a_3^{(HM)} \zeta^3 + a_4^{(HM)} \zeta^4 + \mathcal{O}(\zeta^5),
    \end{align}
where
    \begin{align}
    & a_2^{(HM)} \equiv 8 (1+X) \tilde{z}_-, \qquad a_3^{(HM)} = - \frac{94041 + a_{3}^{(1)} + a_{3}^{(2)}}{(9 + 8 X)^5}, \qquad a_4^{(HM)} = 16 (1 + X) \frac{9477 + a_{4}^{(1)} + a_{4}^{(2)}}{(9 + 8 X)^{6} \, \tilde{z}_-},\nn\\
    &  a_{3}^{(1)} =  128 (X (X+1))^{3/2} (2 X+1) (8 X+9)^2,\nn\\
    & a_{3}^{(2)} = 8 X (4 X (8 X (X (16 X (4 X+41)+2345)+4104)+30753)+59535),\nn\\
    & a_{4}^{(1)} = 8 \sqrt{X^3 (X+1)} (8 X (4 X+1) (8 X+9)-27), \nn\\
    &a_{4}^{(2)} = 4 X (2 X (8 X (4 X (8 X+79)+859)+8397)+9963).
\end{align}
 In Fig. \ref{fig: a2-2-p1-sound} the conditions $a^{(HM)}_n - n$ are plotted against $X$ for $n = 2, 3, 4$, represented by the corresponding colors. Checking the Bieberbach conjecture specified by $a_n \leq n$, will give rise to the condition $X \leq 0.853$ and ensures all plots remain below the zero line. We restrict ourselves to $z>0$ and therefore the LM region has nothing to do with it.

\begin{figure}
    \centering
    \includegraphics[width=0.475\textwidth, valign=t]{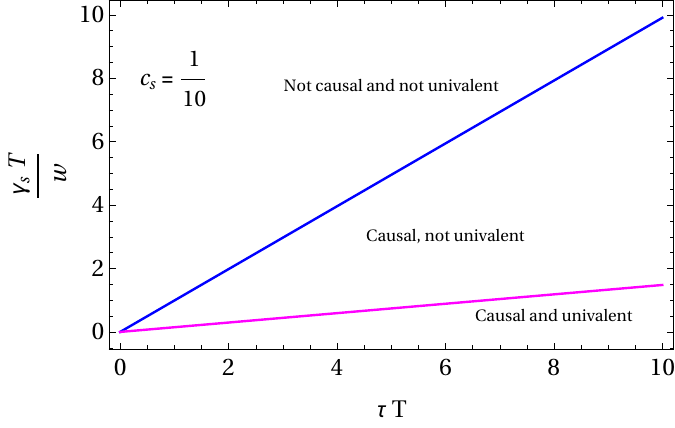}
    \hspace{0.2cm}
    \includegraphics[width=0.475\textwidth, valign=t]{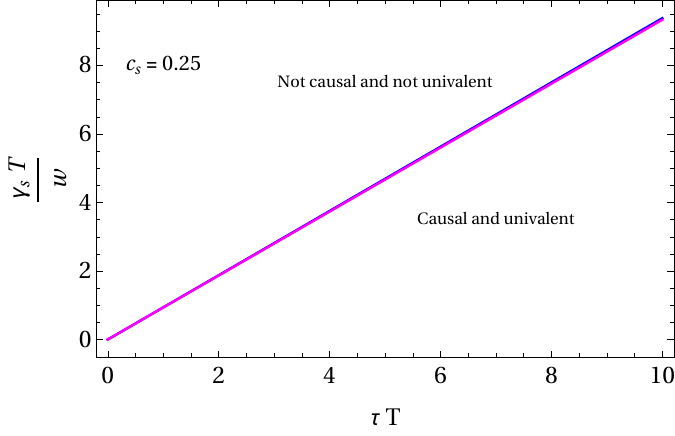}\\
    \vspace{0.5cm}
    \includegraphics[width=0.5\textwidth,  valign=t]{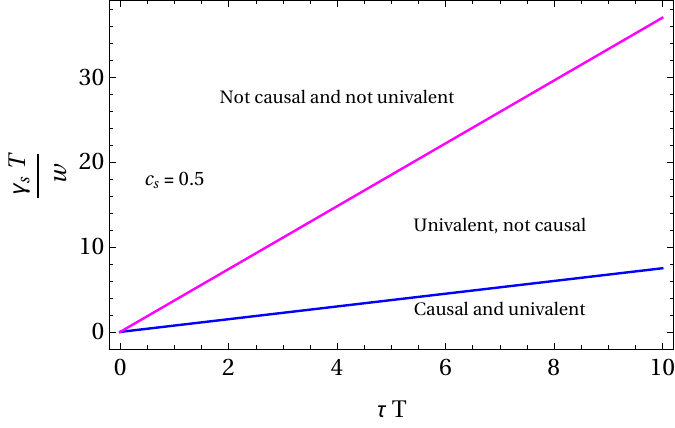}
    \caption{Acceptable region for transports $(\gamma_s, \tau)$ by combining the causality condition in Eq. \eqref{eqcausal} and univalence bounds in Eq. \eqref{eqb}. In each plot, the blue (magenta) lines denote the boundary of causality (univalence) conditions that in below (above) which the conditions are satisfied (violated). Each plot is produced by a certain $c_s$ which is shown inside.}
    \label{fig-univ-causal}
\end{figure}

In Fig. \ref{fig-univ-causal} we combine the univalence and causality conditions to obtain the space in which all constraints are satisfied. Stability is assumed on each point and the $c_s$ value by which plots are produced is shown inside. The blue line denotes the causality boundary below (above) which the causality preserves (violates). The magenta line defines the univalence boundary below (above) which the univalence preserves (violates). At small $c_s$, the causality bound lies above the univalence and there is a region where the theory is causal but is not univalent. At $c_s = 0.25$, these two bounds meet, and for $c_s \geq 0.25$ the univalence is above the causality. In this case, there is a univalent region in the middle, but causality is missed.
\section{Interplay of bounds in the BDNK model}
We employ the BDNK model for an uncharged and conformal fluid, which has constitutive relations given in Eq. \eqref{eqtj}. The relations involve transport coefficients that absorb frame change effects, similar to gauge parameters
\begin{align}
    & \mathcal{E} = \varepsilon + \epsilon_1 \frac{u^\mu \partial_\mu T}{T} + \epsilon_2 \partial_\mu u^\mu + \mathcal{O}(\partial^2),\nn\\
    & \mathcal{P} = p + \pi_1 \frac{u^\mu \partial_\mu T}{T} + \pi_2 \partial_\mu u^\mu + \mathcal{O}(\partial^2),\nn\\
    & \mathcal{Q}^\mu = \theta_1 u^\nu \partial_\nu u^\mu + \theta_2 \frac{\Delta^{\mu \nu} \partial_\nu T}{T} + \mathcal{O}(\partial^2),\nn\\
    &\mathcal{T}^{\mu \nu} = - 2 \eta \sigma^{\mu \nu} + \mathcal{O}(\partial^2),\nn\\
    & \mathcal{N} = 0, \qquad \mathcal{J}^\mu
    = 0.
 \end{align}
 In the conformal limit and under thermodynamic consistency conditions, the transport coefficients are related as \cite{Kovtun:2019hdm}
 \begin{align}
     &\pi_2 = c_s^2 \left(\epsilon_2  + \pi_1 - c_s^2 \epsilon_1\right),\nn\\
     &\theta_1 = \theta_2.
 \end{align}
 This results in five transports $(\epsilon_1, \epsilon_2, \pi_1, \theta_1, \eta)$ to determine the fluid's behavior in the first order of derivatives. Eq.  \eqref{eqce} yields the following result for small hydro perturbations in the shear and sound channels
 \begin{align}
     \mbox{Shear:} \qquad & \omega^2 + i \frac{w \omega}{\theta} - \frac{\eta z}{\theta} = 0,\\
     \mbox{Sound:} \qquad & c_{s}^2 \epsilon_{1} \theta \omega^4 +i w \left(c_{s}^2 \epsilon_{1}+\theta\right)\omega^3 -i z w \left(\gamma_{s}+c_{s}^4\epsilon_{1}+c_{s}^2\theta \right) \omega \nn\\
       &-\bigg(w^2+z c_{s}^2 (c_{s}^4\epsilon_{1}^2+\gamma_{s}\epsilon_{1}+(\epsilon_{2}+\pi_{1})(\theta-c_{s}^2\epsilon_{1})+\epsilon_{2}\pi_{1})\bigg)\omega^2\nn\\
       &+z c_{s}^2\left(w^2 + z \theta (c_{s}^2(\epsilon_{2}+\pi_{1}-c_{s}^2 \epsilon_{1})-\gamma_{s})\right) = 0.
 \end{align}
 Hereafter, we take $\theta = \theta_1$. In the following subsections, we study the analytical properties of these channels separately.
\subsection{Shear channel}
For the shear channel, the determinant equation is given by
\begin{align}\label{eq20}
    F^{\text{BDNK}}_\text{shear}(\omega, k^2) = \omega^2 + i \frac{w \omega}{\theta} - \frac{\eta z}{\theta} = 0.
\end{align}
The stability and causality conditions are
\begin{align}
    \mbox{Stability}\,\,&, \,\,\, \eta>0, \,\,\, \theta>0,\nn\\
    \mbox{Causality}\,\,&, \,\,\, \eta < \theta.
\end{align}
Likewise, solutions of the imaginary frequency to Eq. \eqref{eq20} are
\begin{align}\label{eq: shear-BDNK}
    \beta_\pm(z) = \frac{w}{2 \theta} \left(-1 \pm \sqrt{1 -  \frac{4 \eta z \theta}{w^2}}\right).
\end{align}
The point $z_c = \frac{w^2}{4 \eta \, \theta}$ is the collision point, and the modes' plot is similar to Fig. \ref{fig1}. The univalence bound goes in the same way as the MIS model and gives no new conditions on transports. 
\subsection{Sound channel}
In the sound channel identified by the equation for longitudinal perturbations, the determining equation is given by\cite{Kovtun:2019hdm}
\begin{align}\label{eq: disp-sc}
  F_\text{sound}^\text{BDNK} (\omega, z) = & c_{s}^2 \epsilon_{1} \theta \omega^4 +i w \left(c_{s}^2 \epsilon_{1} + \theta \right)\omega^3 -i z w \left(\gamma_{s}+c_{s}^4\epsilon_{1}+c_{s}^2\theta\right) \omega \nn\\
       &-\bigg(w^2+z c_{s}^2 (c_{s}^4\epsilon_{1}^2+\gamma_{s}\epsilon_{1}+(\epsilon_{2}+\pi_{1})(\theta-c_{s}^2\epsilon_{1})+\epsilon_{2}\pi_{1})\bigg)\omega^2\nn\\
       &+z c_{s}^2\left(w^2 + z \theta (c_{s}^2(\epsilon_{2}+\pi_{1}-c_{s}^2 \epsilon_{1})-\gamma_{s})\right) = 0.
\end{align}
The Routh-Hurwitz stability criteria provide us with the following conditions
\begin{align}\label{eq: sc-stab}
    &\epsilon_1 > 0, \quad \bar{\theta} > 0, \quad \eta > 0, \quad -1 + c_s^2 \left( \bar{\epsilon}_2 - \bar{\epsilon}_1 + \bar{\pi}_1 \right) > 0,\\
    & \frac{\bar{\epsilon_{1}}^2}{c_{s}^2}+c_{s}^2(\bar{\epsilon_{1}}-\bar{\epsilon_{2}})(\bar{\epsilon_{1}}+\bar{\theta})^2(\bar{\epsilon_{1}}-\bar{\pi_{1}})+(\bar{\epsilon_{1}}+\bar{\theta})(2\bar{\epsilon_{1}}^2-\bar{\epsilon_{1}}(\bar{\epsilon_{2}}+\bar{\pi_{1}})+(\bar{\theta}+\bar{\epsilon_{2}})(\bar{\theta}+\bar{\pi_{1}}))>0,\nn
\end{align}
where the bar notation represents the dimensionless quantities as follows
\begin{align}\label{eq: dimless-t-BDNK}
    \bar{\epsilon}_1 = \frac{c_s^2 \epsilon_1}{\gamma_s}, \quad \bar{\epsilon}_2 = \frac{\epsilon_2}{\gamma_s}, \quad \bar{\theta} = \frac{\theta}{\gamma_s}, \quad \bar{\pi}_1 = \frac{\pi_1}{\gamma_s}.
\end{align}
The causality condition comes from the large $k$ solutions of the Eq. \eqref{eq: disp-sc}. Using the ansatz $\omega = v_{sound} k_z$ in large $k_z$ and keeping only the dominant terms, the following constraint is obtained for the ultimate speed of fluctuations' propagation
\begin{align}\label{eq: sc-caus}
    & v_{sound} = \lim_{k \to \infty} \bigg|\frac{\partial Re (\omega)}{\partial k}\bigg| = \sqrt{\frac{\mathcal{S}_1 + \mathcal{S}_2 + \sqrt{4 c_s^2 \bar{\theta}^2 \mathcal{S}_1 + (\mathcal{S}_1 + \mathcal{S}_2)^2}}{2 \bar{\epsilon}_1 \bar{\theta}}}\leq 1,
    \end{align}
where
    \begin{align}
    & \mathcal{S}_1 \equiv \bar{\epsilon}_1 \left(1 + c_s^2 \left( \bar{\epsilon}_1 - \bar{\epsilon}_2 - \bar{\pi}_1 \right)\right), \quad \mathcal{S}_2 = c_s^2 \left(\bar{\theta} \bar{\pi}_1 + \bar{\epsilon}_2 (\bar{\theta} + \bar{\pi}_1)\right).
\end{align}
To determine the physical zone of transports, the conditions \eqref{eq: sc-stab} and \eqref{eq: sc-caus} have to be considered simultaneously and their common region is identified as the allowable region. The solutions are depicted in the right part of Fig. 3 in \cite{Kovtun:2019hdm} with the choices $(\bar{\epsilon}_2 = 0, \bar{\pi}_1 = 3/c_s^2)$ and for different $c_s$.

\begin{figure}
    \centering
    \includegraphics[width=0.45\textwidth, valign=t]{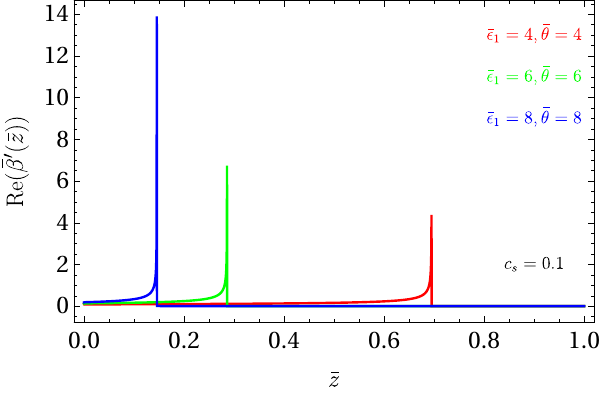}
    \hspace{0.4cm}
    \includegraphics[width=0.45\textwidth, valign=t]{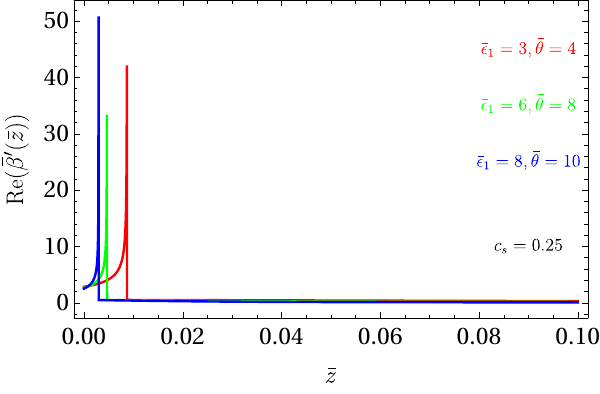}
    \\
    \includegraphics[width=0.45\textwidth, valign=t]{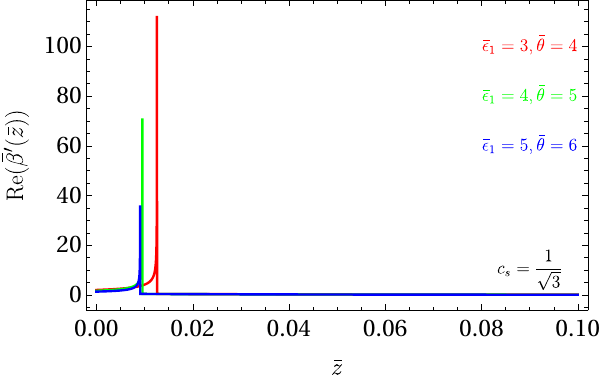}
    \hspace{0.4cm}
    \includegraphics[width=0.45\textwidth, valign=t]{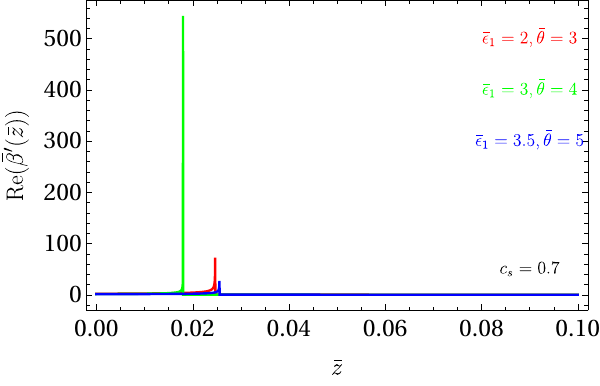}
    \caption{Plot $Re (\bar{\beta}'(\bar{z})) $ in terms of $\bar{z}$ with $\bar{\epsilon}_2 = 0$, and $ \bar{\pi}_1 = 3/c_s^2$. The top left (right) panel corresponds to $c_s = 0.1 (c_s = 0.25)$, and the bottom left (right) panel addresses $c_s = 1/\sqrt{3} (c_s = 0.7)$. Identified colors stand for specific choices of $(\bar{\epsilon}_1, \bar{\theta})$ as shown.}
    \label{fig: betap-BDNK}
\end{figure}

To derive the univalence bounds, it is essential to verify $Re (\bar{\beta}'(z)) > 0$.  After replacing $\omega = i \beta(z)$ in Eq. \eqref{eq: disp-sc}, we obtain
\begin{align}\label{eq: betap-BDNK}
    &\bar{\beta}'(\bar{z}) = \frac{\partial \bar{\beta}(\bar{z})}{\partial \bar{z}}= \frac{\boldsymbol{n}_0 + \boldsymbol{n}_1 \bar{\beta}(\bar{z}) + \boldsymbol{n}_2 \bar{\beta}(\bar{z})^2}{\boldsymbol{d}_0 + \boldsymbol{d}_1 \bar{\beta}(\bar{z}) + \boldsymbol{d}_2 \bar{\beta}(\bar{z})^2 + \boldsymbol{d}_3 \bar{\beta}(\bar{z})^3},
    \end{align}
with the following definitions
    \begin{align}
    &\boldsymbol{n}_0 = c_s^2 \left(2 \bar{\theta}  \, \bar{z} \left(c_s^2 (-\bar{\pi}_1+\bar{\epsilon}_1-\bar{\epsilon}_2)+1\right)-1\right) ,\nn\\
    &\boldsymbol{n_1} = -\left(c_s^2 (\bar{\theta} +\bar{\epsilon}_1)+1\right),\nn\\
    &\boldsymbol{n_2} = -\left(c_s^2 \bar{\epsilon}_1^2-c_s^2 \bar{\epsilon}_1 (\bar{\pi}_1+\bar{\epsilon}_2)+c_s^2 (\bar{\theta}  \bar{\pi}_1+\bar{\epsilon}_2 (\bar{\theta} +\bar{\pi}_1))+\bar{\epsilon}_1\right),\nn\\
    &\boldsymbol{d_0} =   \bar{z} \left(c_s^2 (\bar{\theta} +\bar{\epsilon}_1)+1\right), \nn\\
    &\boldsymbol{d}_1 = 2 \left(\bar{z} \left(c_s^2 \bar{\epsilon}_1^2-c_s^2 \bar{\epsilon}_1 (\bar{\pi}_1+\bar{\epsilon}_2)+c_s^2 (\bar{\theta}  \bar{\pi}_1+\bar{\epsilon}_2 (\bar{\theta} +\bar{\pi}_1))+\bar{\epsilon}_1\right)+1\right),\nn\\
    & \boldsymbol{d}_2 = 3 (\bar{\theta} +\bar{\epsilon}_1),\qquad \boldsymbol{d}_3 = 4 \bar{\theta}  \, \bar{\epsilon}_1.
\end{align}
and the bar notations define dimensionless quantities 
\begin{align}\label{eq: dimless-oz-BDNK}
    \bar{\beta}  =  \beta \frac{\gamma_s}{w}, \qquad \bar{z} = z\frac{\gamma_s^2}{w^2}.
\end{align}
To evaluate Eq. \eqref{eq: betap-BDNK} using the solutions of Eq. \eqref{eq: disp-sc}, we must first rewrite the latter in terms of dimensionless quantities. By applying the replacements from Eqs. \eqref{eq: dimless-t-BDNK} and \eqref{eq: dimless-oz-BDNK}, the solutions can be expressed as
\begin{align}\label{eq: on-shell-4}
    (\bar{\beta}_\star(\bar{z}))_{s_1}^{s_2} =& -\frac{\bar{a}_3}{4 \bar{a}_4} + s_1 \frac{1}{2}\bigg( \frac{\bar{a}_3^2}{4 \bar{a}_4^2} - \frac{2 \bar{a}_2}{3 \bar{a}_4} + \frac{2^{1/3} \mathcal{C}}{3 \bar{a}_4 \left(\mathcal{D} + \mathcal{G}\right)^{1/3}} + \frac{\left(\mathcal{D} + \mathcal{G}\right)^{1/3}}{ 2^{1/3} \times 3 \bar{a}_4}\bigg)^{1/2} \nn\\
    & + s_2 \frac{1}{2} \bigg( \frac{\bar{a}_3^2}{2 \bar{a}_4^2} - \frac{4 \bar{a}_2}{3 \bar{a}_4} - \frac{2^{1/3} \mathcal{C}}{3 \bar{a}_4 \left(\mathcal{D} + \mathcal{G}\right)^{1/3}} - \frac{\left(\mathcal{D} + \mathcal{G}\right)^{1/3}}{ 2^{1/3} \times 3 \bar{a}_4}- \frac{\mathcal{F}}{4 \bigg(\frac{\bar{a}_3^2}{4 \bar{a}_4^2} - \frac{2 \bar{a}_2}{3 \bar{a}_4} + \frac{2^{1/3} \mathcal{C}}{3 \bar{a}_4 \left(\mathcal{D} + \mathcal{G}\right)^{1/3}}\bigg)^{1/2}}\bigg)^{1/2},
\end{align}
where
\begin{align}\label{eq: coeff-BDNK}
    &\mathcal{C} \equiv 12 \bar{a}_0 \bar{a}_4 - 3 \bar{a}_1 \bar{a}_3 + \bar{a}_2^2, \nn\\
    &\mathcal{D} \equiv -72 \bar{a}_0 \bar{a}_2 \bar{a}_4 + 27 \bar{a}_0 \bar{a}_3^2 + 27 \bar{a}_1^2 \bar{a}_4 - 9 \bar{a}_1 \bar{a}_2 \bar{a}_3 + 2 \bar{a}_2^3,\nn\\
    &\mathcal{G}^2 \equiv \mathcal{D}^2 -4 \mathcal{C}^3, \nn\\
    &\mathcal{F} \equiv -\frac{8 \bar{a}_1}{\bar{a}_4} + \frac{4 \bar{a}_2 \bar{a}_3}{\bar{a}_4^2} - \frac{\bar{a}_3^3}{\bar{a}_4^3}.
\end{align}
and $(s_1, s_2)$ take either $\pm$. The $\bar{a}_i$ denote the coefficients of $\bar{\beta}^i$ for $i = 0, 1, 2, 3, 4$ in the Eq.  \eqref{eq: disp-sc}, and they are provided as follows
\begin{align}\label{eq: coeff-dimless-BDNK}
    &\bar{a}_4 \equiv  \bar{\epsilon}_1 \bar{\theta}, \nn\\
    &\bar{a}_3 \equiv  \bar{\epsilon}_1 + \bar{\theta}, \nn\\
    &\bar{a}_2 \equiv 1 + \bar{z} \bigg(c_s^2 \bar{\epsilon}_1 (\bar{\epsilon}_1 - \bar{\pi}_1 - \bar{\epsilon}_2)+c_s^2 (\bar{\theta}  \bar{\pi}_1+\bar{\epsilon}_2 (\bar{\theta} +\bar{\pi}_1))+\bar{\epsilon}_1\bigg),\nonumber\\
    &\bar{a}_1 \equiv \bar{z} \left(1 + c_{s}^2(\bar{\epsilon}_{1} + \bar{\theta})\right), \nn\\
    &\bar{a}_0 \equiv c_s^2 \bar{z} \left(\bar{\theta}  \bar{z} \left(c_s^2 (\bar{\pi}_1-\bar{\epsilon}_1+\bar{\epsilon}_2)-1\right)+1\right).
\end{align}
We derived the real part of the derivative, $Re(\bar{\beta}'(\bar{z}))$, using the solutions given by Eq. \eqref{eq: on-shell-4}. We noticed that the condition $Re(\bar{\beta}'(\bar{z})) > 0$ holds only for $(\bar{\beta}\star)_-^-$. In Fig. \ref{fig: betap-BDNK}, we present the $Re(\bar{\beta}'(\bar{z}))$ plot for different values of $c_s = (0.1, 0.25, \frac{1}{\sqrt{3}}, 0.7)$, each corresponding to $(\bar{\epsilon}_2 = 0, \bar{\pi}_1 = 3/c_s^2)$. The identified colors represent distinct choices of $(\bar{\epsilon}_1, \bar{\theta})$. These choices are made to ensure they belong to the stable and causal regions depicted in Fig. 3 of \cite{Kovtun:2019hdm}. The observed bumps in Fig. \ref{fig: betap-BDNK} correspond to the locations where the denominator of $\bar{\beta}(\bar{z})$ in Eq. \eqref{eq: betap-BDNK} becomes zero.

To derive the univalent series of $(\bar{\beta}_\star(\bar{z}))_-^- = (\bar{\beta}_\star(\phi^{-1}(\xi)))_-^-$, we first require the appropriate conformal map based on the analytical solutions of Eq. \eqref{eq: disp-sc}, as presented in Eq. \eqref{eq: on-shell-4}. This map must contain information about the singularity structures originating from the absolute values of the roots of $\mathcal{G}(\bar{z}) = 0$ in Eq. \eqref{eq: coeff-BDNK}
\begin{align}\label{eq: ebar}
    0 = \mathcal{G}(\bar{z}) = \sum\limits_{n=0}^6 \bar{g}_n \bar{z}^n,
\end{align}
where the $\bar{g}_n$ coefficients are expressed in Appendix \ref{sec: App-A}. Generally, Eq. \eqref{eq: ebar} has six different roots but the root numbers may vary depending on the specific transports. Additionally, one root is always $\bar{z} = 0$ due to $\bar{g}_0 = 0$. Since there is no analytical solution for Eq. \eqref{eq: ebar}, we resort to a numerical approach to solve it for given transports. Consequently, the procedure to obtain the map and univalent series is carried out point by point in the transport space $(\bar{\epsilon}_1, \bar{\theta})$.

\begin{figure}
\centering
\begin{tikzpicture}[scale=0.9]
\draw[->] (-4,0) -- (4,0) node[below, yshift = -0.10cm, scale = 1.25]{\textbf{Re $\bar{z}$}};
\draw [->](0,-4) -- (0,4) node[above,  xshift = -0.65cm, yshift = -0.45cm, scale = 1.25]{\textbf{Im $\bar{z}$}};
\draw [black, thick](1,-4) -- (1,4);
\draw  [red, thick, arrows = {-Stealth[]}](1.25,1) -- (3.75,1) 
node[above, yshift = 0.15cm, scale = 1.2]{HM};
\draw  [black, dashed, ultra thick] (2,0) -- (4,0);
\draw [blue, thick, arrows = {-Stealth[]}](0.75 ,1) -- (-2,1) 
node[above, yshift = 0.15cm, scale = 1.2]{LM};
\draw  [black, dashed, ultra thick] (0,0) -- (1,0);
\filldraw [black] (0,0) circle (3pt) node[below, xshift = -0.2cm, yshift = -0.02cm, scale = 1.25]{$\bar{z}_0$};
\filldraw [black] (1,0) circle (3pt) 
node[below, xshift = -0.2cm, yshift = -0.02cm, scale = 1.25]{$\bar{z}_1$};
\filldraw [black] (2,0) circle (3pt) node[below , yshift = -0.02cm, scale = 1.25]{$\bar{z}_2$};
\end{tikzpicture}
\hspace{0.4cm}
\begin{tikzpicture}[scale=0.9]
\draw [->](-4,0) -- (5,0) node[below, yshift = -0.10cm, scale = 1.25]{\textbf{Re $\bar{z}$}};
\draw [->](0,-4) -- (0,4) node[above,  xshift = -0.65cm, yshift = -0.45cm, scale = 1.25]{\textbf{Im $\bar{z}$}};
\filldraw [black] (0,0) circle (3pt) node[below, xshift = -0.25cm, yshift = -0.05cm, scale = 1.25]{$\bar{z}_0$};
\filldraw [black] (1,0) circle (3pt) node[below, xshift = -0.05cm, yshift = -0.05cm, scale = 1.25]{$\bar{z}_1$};
\filldraw [black] (2,0) circle (3pt) node[below, xshift = -0.2cm, yshift = -0.05cm, scale = 1.25]{$\bar{z}_2$};
\filldraw [black] (3,0) circle (3pt) node[below, xshift = -0.05cm, yshift = -0.05cm, scale = 1.25]{$\bar{z}_3$};
\draw [black, thick](2,-4) -- (2,4);
\draw  [red, thick,arrows = {-Stealth[]}](2.3,1) -- (5,1) node[above, yshift = 0.15cm, scale=1.3]{HM};
\draw  [black, dashed, ultra thick] (3,0) -- (5,0);
\draw [blue, thick,arrows = {-Stealth[]}](1.85,1) -- (0,1) node[above, yshift = 0.15cm, xshift = 0.45cm, scale=1.3]{LM};
\draw  [black, dashed, ultra thick] (1,0) -- (2,0);
\draw  [black, dashed, ultra thick] (-4,0) -- (0,0) node[above, yshift = 0.15cm, xshift= -2.05cm, scale=1.1]{Not acceptable region};
\end{tikzpicture}
\begin{tikzpicture}[scale=0.65]
\draw [->](-5,0) -- (6,0) node[below, yshift = -0.10cm, scale = 1.25]{\textbf{Re $\bar{z}$}};
\draw [->] (0,-6) -- (0,6) node[below, yshift = -0.05cm, xshift= -0.65cm, scale = 1.25]{\textbf{Im $\bar{z}$}};
\filldraw [black] (0,0) circle (3pt) node[below, xshift = -0.2cm, yshift = -0.1cm, scale = 1.25]{$\bar{z}_0$};
\filldraw [black] (1,0) circle (3pt) node[below, xshift = -0.2cm, yshift = -0.1cm, scale = 1.25]{$\bar{z}_1$};
\filldraw [black] (2,0) circle (3pt) node[below, xshift = -0.1cm, yshift = -0.1cm, scale = 1.25]{$\bar{z}_2$};
\filldraw [black] (3,0) circle (3pt) node[below, xshift = -0.2cm, yshift = -0.1cm, scale = 1.25]{$\bar{z}_3$};
\filldraw [black] (4,0) circle (3pt) node[below, xshift = -0.1cm, yshift = -0.1cm, scale = 1.25]{$\bar{z}_4$};
\draw [black, thick](1,-6) -- (1,6);
\draw [black, thick](3,-6) -- (3,6);
\draw  [red, ultra thick,arrows = {-Stealth[]}](3.5,1) -- (5.5,1) node[above, yshift = 0.1cm, xshift = 0.1cm, scale=1.3]{HM};
\draw  [black, dashed, ultra thick] (4,0) -- (6,0);
\draw  [blue, ultra thick,arrows = {-Stealth[]}](0.8,1) -- (-3,1) node[above, yshift = 0.15cm, xshift = 0.45cm, scale=1.3]{LM};
\draw  [black, dashed, ultra thick] (0,0) -- (1,0);
\draw [<->] [green, ultra thick] (1,1) -- (3,1) node[above, yshift = 0.15cm, xshift = -0.75cm, scale=1.3]{IM};
\draw  [black, dashed, ultra thick] (2,0) -- (3,0);
\end{tikzpicture}
\hspace{0.15cm}
\begin{tikzpicture}[scale=0.65]
\draw [->](-5,0) -- (7,0) node[below, yshift = -0.10cm, scale = 1.25]{\textbf{Re $\bar{z}$}};;
\draw [->](0,-6) -- (0,6) node[below, yshift = -0.10cm, xshift= -0.65cm, scale = 1.25]{\textbf{Im $\bar{z}$}};;
\filldraw [black] (0,0) circle (3pt) node[below, xshift = -0.2cm, yshift = -0.05cm, scale = 1.25]{$\bar{z}_0$};
\filldraw [black] (1,0) circle (3pt) node[below, xshift = -0.05cm, yshift = -0.05cm, scale = 1.25]{$\bar{z}_1$};
\filldraw [black] (2,0) circle (3pt) node[below, xshift = -0.15cm, yshift = -0.05cm, scale = 1.25]{$\bar{z}_2$};
\filldraw [black] (3,0) circle (3pt) node[below, xshift = -0.1cm, yshift = -0.05cm, scale = 1.25]{$\bar{z}_3$};
\filldraw [black] (4,0) circle (3pt) node[below, xshift = -0.2cm, yshift = -0.05cm, scale = 1.25]{$\bar{z}_4$};
\filldraw [black] (5,0) circle (3pt) node[below, xshift = -0.05cm, yshift = -0.05cm, scale = 1.25]{$\bar{z}_5$};
\draw [black, thick](2,-6) -- (2,6);
\draw [black, thick](4,-6) -- (4,6);
\draw  [red, ultra thick,arrows = {-Stealth[]}](4.25,1) -- (6.85,1) node[above, yshift = 0.15cm, xshift = -0.45cm, scale=1.3]{HM};
\draw  [black, dashed, ultra thick] (5,0) -- (7,0);
\draw  [blue, ultra thick,arrows = {-Stealth[]}](1.85,1) -- (0,1) node[above, yshift = 0.15cm, xshift = 0.55cm, scale=1.3]{LM};
\draw  [black, dashed, ultra thick] (1,0) -- (2,0);
\draw  [<->][green, ultra thick] (2,1) -- (4,1) node[above, yshift = 0.15cm, xshift = -0.75cm, scale=1.3]{IM};
\draw  [black, dashed, ultra thick] (3,0) -- (4,0);
\draw  [black, dashed, ultra thick] (-5,0) -- (0,0) node[above, yshift = 0.15cm, xshift= -1.95cm, scale=1.2]{Not acceptable region};
\end{tikzpicture}
\caption{Analytical structure of solutions for dispersion relation of BDNK model in the sound channel. The dashed lines on the horizontal lines refer to the branch cuts and vertical solid lines separate them into the HM (High Momentum), IM (Intermediate Momentum), and LM (Low Momentum) regions. The black dots stand for the absolute values of roots of Eq. \eqref{eq: ebar}. Different momentum zones are separated by vertical solid lines.}\label{fig: sol-BDNK}
\end{figure}
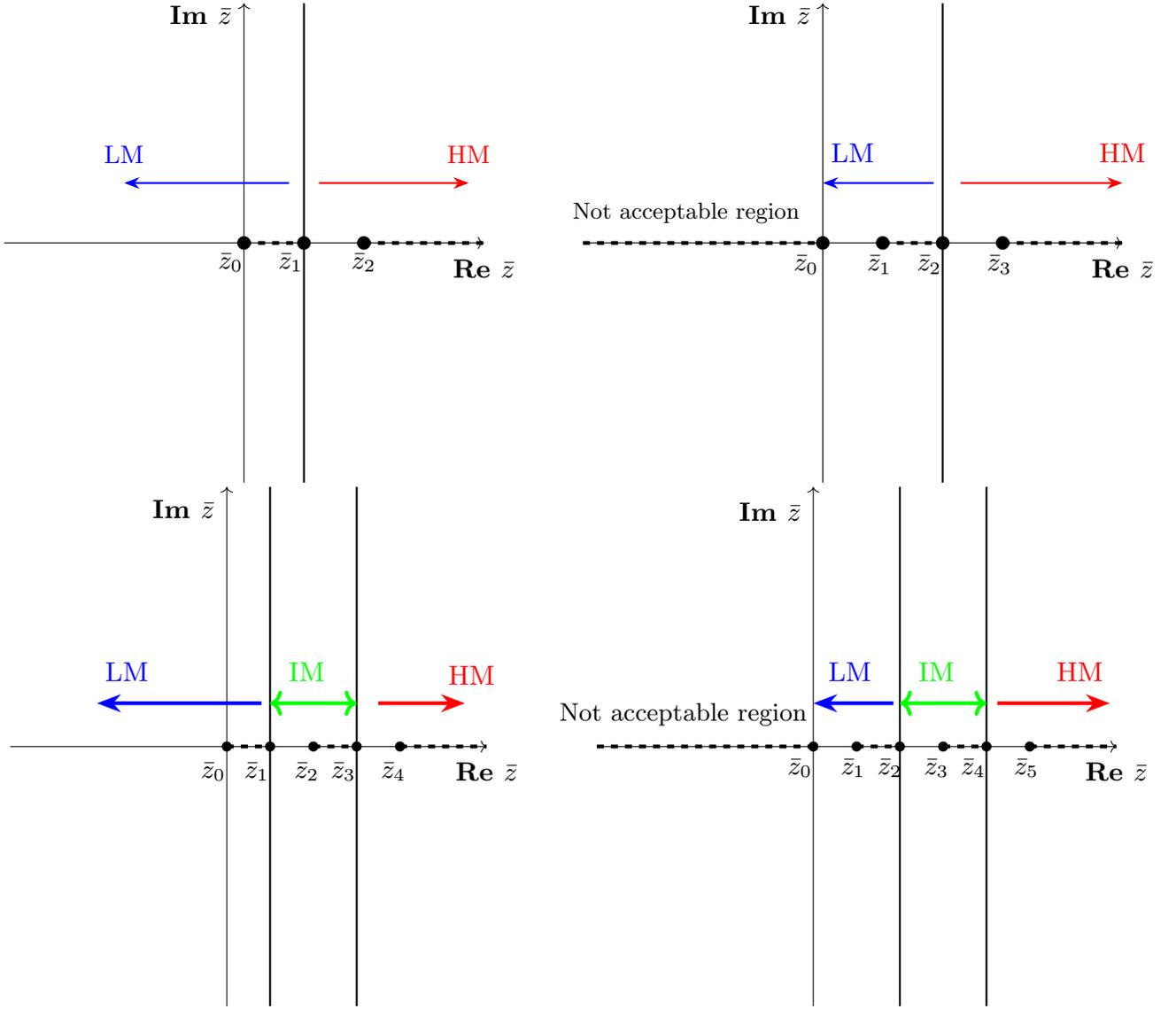

It is crucial to note that the desired conformal map's properties and behavior are highly dependent on the number of roots, as the branch cut locations differ for each case. In Fig. \ref{fig: sol-BDNK}, we illustrate the branch cut structures, which vary with the number of roots. These structures are divided into HM (High Momentum), IM (Intermediate Momentum), and LM (Low Momentum) zones separated by vertical solid lines. This partitioning is essential because constructing a global map for each set is not feasible. In Table \ref{table: expansion}, we explain the possibilities of having different types of expansions. HME (High Momentum Expansion) and LME (Low Momentum Expansion) exist for all root number scenarios, while IME (Intermediate Momentum Expansion) only appears when the root numbers are $n = 5$ and $n = 6$. We disregard regions where the branch cut lies on $Re(\bar{z}) \leq 0$, as in such cases, that part does not represent an analytic section.

For the HM region (the red parts in Fig. \ref{fig: sol-BDNK}) we employ the following map
\begin{align}
    -\frac{\bar{z}}{4(\bar{z}-\bar{z}_\text{Max})} = \frac{\zeta}{(1-\zeta)^2}, \,\,\, \Rightarrow \quad \bar{z} = \phi_{(HM)}^{-1}(\zeta) = \frac{4 \bar{z}_\text{Max} \zeta}{(1+\zeta)^2},
\end{align}
where $\bar{z}_\text{Max}$ is the largest root in each plot. For the IM region (the green parts in Fig. \ref{fig: sol-BDNK}) which functions at $n=5$ or $n=6$, we benefit from the following map
\begin{align}
    -\frac{\bar{z}_{n-1}-\bar{z}_{n-2}}{4 (\bar{z}-\bar{z}_{n-2})}=\frac{\xi }{(1-\xi )^2}, \,\,\, \Rightarrow \quad \bar{z} = \phi_{(IM)}^{-1}(\zeta) = \frac{\bar{z}_{n-2}(\xi +1)^2 - \bar{z}_{n-1}(\xi -1)^2 }{4 \xi }.
\end{align}
For the LM region (the blue parts in Fig. \ref{fig: sol-BDNK}) with $n=3$ or $n=5$ root numbers, we impose the following map
\begin{align}
    -\frac{\bar{z}_1}{4 z}=\frac{\xi }{(1-\xi )^2}, \,\,\, \Rightarrow \quad \bar{z} = \phi_{(LM)}^{-1}(\zeta) = -\frac{(\xi -1)^2 \bar{z}_1}{4 \xi },
\end{align}
 For $n=4$ or $n=6$ root numbers, the LM map is
\begin{align}
    -\frac{\bar{z}_{2}-\bar{z}_{1}}{4 (\bar{z}-\bar{z}_{1})}=\frac{\xi }{(1-\xi )^2}, \,\,\, \Rightarrow \quad \bar{z} = \phi_{(LM)}^{-1}(\zeta) = \frac{\bar{z}_{1}(\xi +1)^2 - \bar{z}_{2}(\xi -1)^2 }{4 \xi }.
\end{align}
In each region, we acquire the conformal map through the following procedure:
\begin{itemize}
\item Initiate by sending the branch cut to the interval $(-\infty, 1/4]$. This can be achieved in two ways for each branch cut: either by moving the head (tail) of the branch cut to $-\infty$ or $1/4$, or vice versa.

\item Next, examine the condition $\vert \xi(\bar{z}) \vert \leq 1$. Determine which map resides within the starting zone of the condition.
\end{itemize}
The chosen map that satisfies this criterion will serve as our desired conformal map for that specific region.
\begin{table}
\centering
\begin{tabular}{|c|c|c|c|c|}
\hline
& $n=3$ & $n=4$ & $n=5$ & $n=6$ \\
\hline
HME & \ding{52} & \ding{52} & \ding{52} & \ding{52} \\
\hline
IME & \ding{56} & \ding{56} & \ding{52} &  \ding{52} \\
\hline
 LME & \ding{52} & \ding{52} & \ding{52} &  \ding{52} \\ 
\hline
\end{tabular}
\caption{Table of allowable HME (High Momentum Expansion), IME (Intermediate Momentum Expansion), and LME (Low Momentum Expansion) for different root numbers, $n$.}\label{table: expansion}
\end{table}
In each momentum zone (HM, IM, and LM), we must insert the corresponding map into the expression
\begin{align}
(\bar{\beta}_\star(\bar{z}))_-^- = (\bar{\beta}_\star(\phi^{-1}(\xi)))_-^- = \xi + \sum\limits_{n=2}^\infty \bar{b}_n \xi^n.
\end{align}
Then, we must verify the Bieberbach conjecture, which states that $\bar{b}_n \leq n$. To do this, we computed the series separately for each region (HM, IM, and LM) up to the $n=100$ terms. We performed these calculations for various points in the $(\bar{\epsilon}_1, \bar{\theta})$ space and different values of $c_s$, with $\bar{\epsilon}_2 = 0$ and $\bar{\pi}_1 = 3/c_s^2$.

\begin{figure}
    \centering
    \includegraphics[width=0.45\textwidth, valign=t]{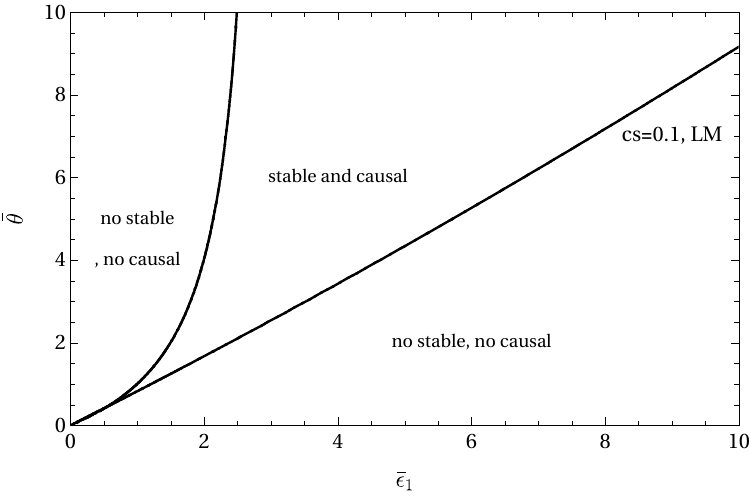}
    \hspace{0.4cm}
    \includegraphics[width=0.45\textwidth, valign=t]{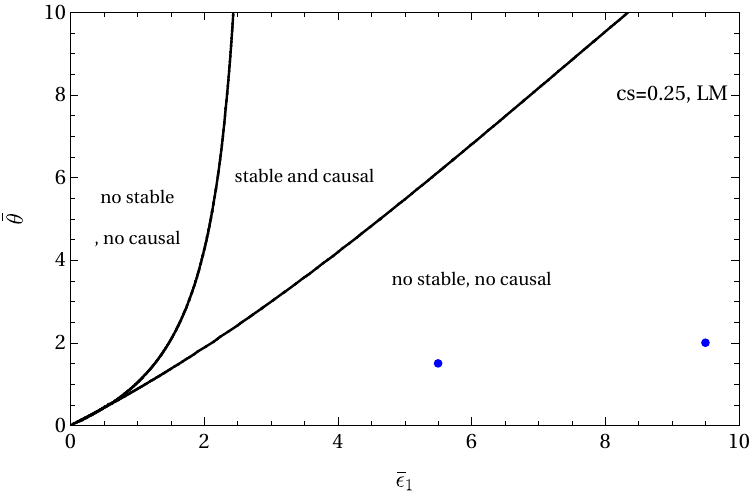}
    \\
    \vspace{0.2cm}
    \includegraphics[width=0.45\textwidth, valign=t]{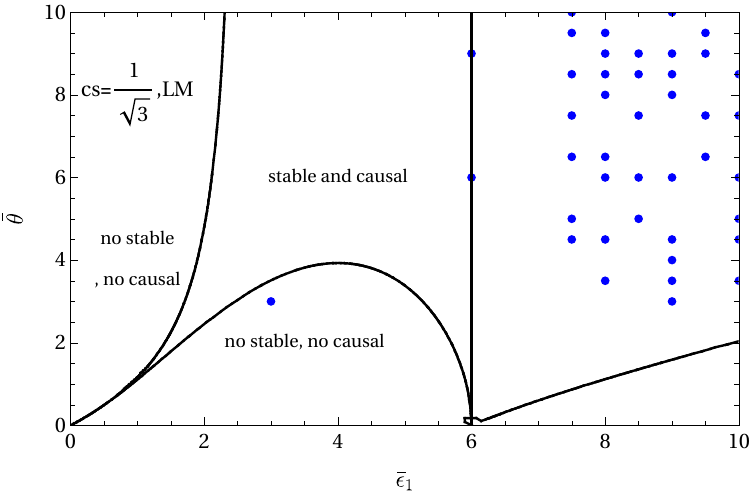}
    \hspace{0.4cm}
    \includegraphics[width=0.45\textwidth, valign=t]{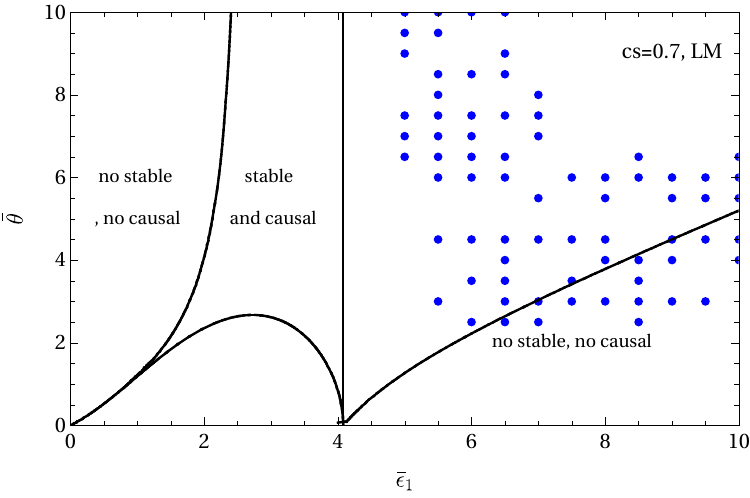}
    \caption{Comparing the stability and causality with univalence in the LM region for the BDNK model with $\bar{\epsilon}_2 = 0$ and $\bar{\pi}_1 = 3/c_s^2$. The top left (right) panels mention the $c_s = 0.1 (c_s = 0.25)$ data, and the bottom left (right) panels refer to the $c_s = 1/\sqrt{3} (c_s = 0.7)$ data. The blue dots represent the points of transport space on which the Bieberbach conjecture is satisfied.}
    \label{fig: LM-BDNK}
\end{figure}

\begin{figure}
    \centering
    \includegraphics[width=0.45\textwidth, valign=t]{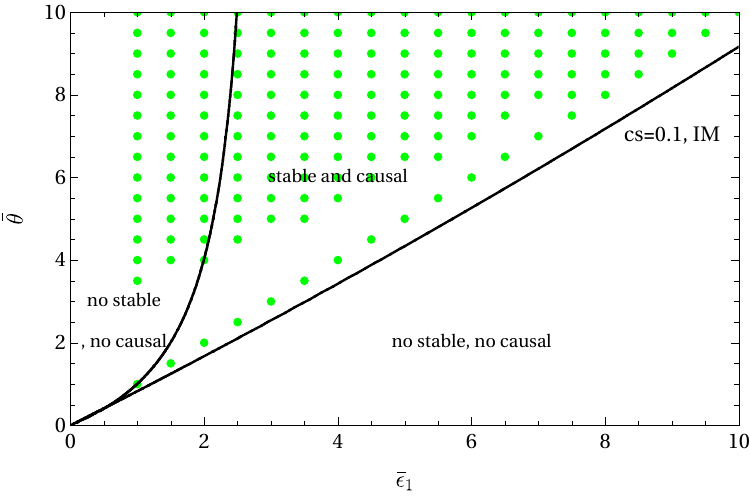}
    \hspace{0.4cm}
    \includegraphics[width=0.45\textwidth, valign=t]{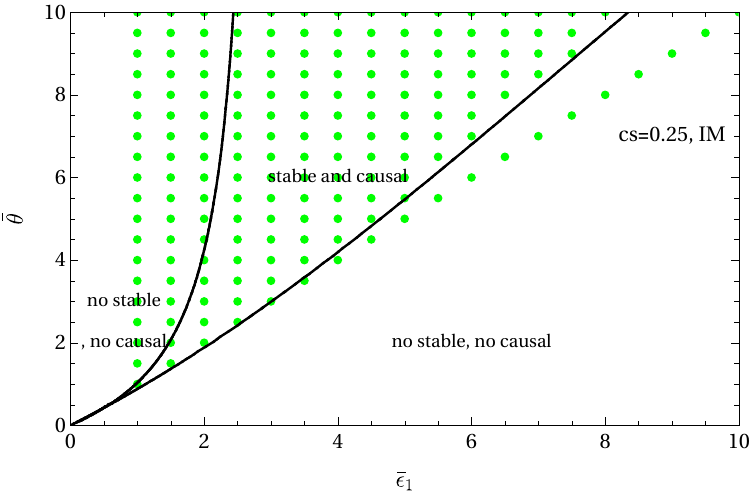}
    \\
    \vspace{0.2cm}
    \includegraphics[width=0.45\textwidth, valign=t]{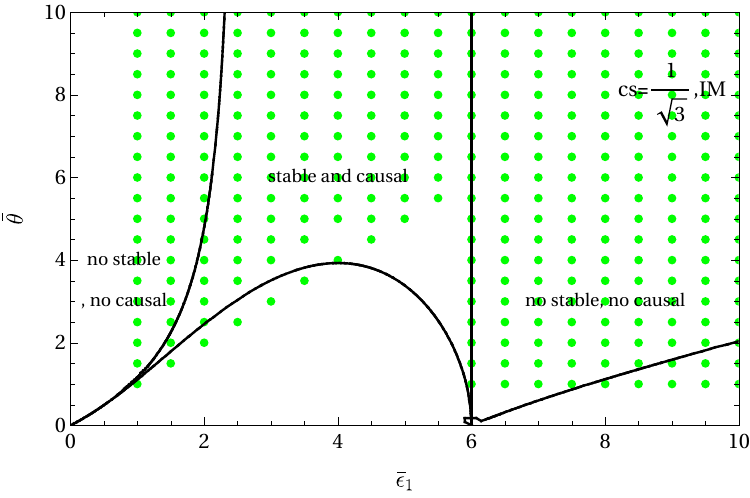}
    \hspace{0.4cm}
    \includegraphics[width=0.45\textwidth, valign=t]{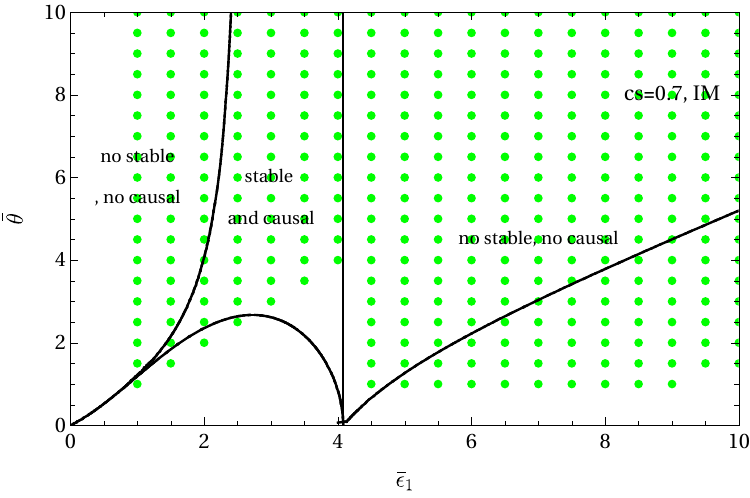}
    \caption{Comparing the stability and causality with univalence in the IM region for the BDNK model with $\bar{\epsilon}_2 = 0$ and $\bar{\pi}_1 = 3/c_s^2$. The top left (right) panels mention the $c_s = 0.1 (c_s = 0.25)$ data, and the bottom left (right) panels refer to the $c_s = 1/\sqrt{3} (c_s = 0.7)$ data. The green dots represent the points of transport space on which the Bieberbach conjecture is satisfied.}
    \label{fig: IM-BDNK}
\end{figure}

\begin{figure}
    \centering
    \includegraphics[width=0.45\textwidth, valign=t]{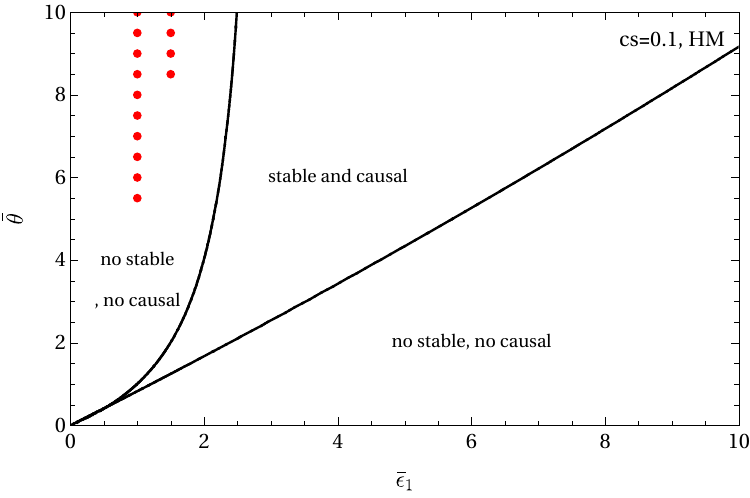}
    \hspace{0.4cm}
    \includegraphics[width=0.45\textwidth, valign=t]{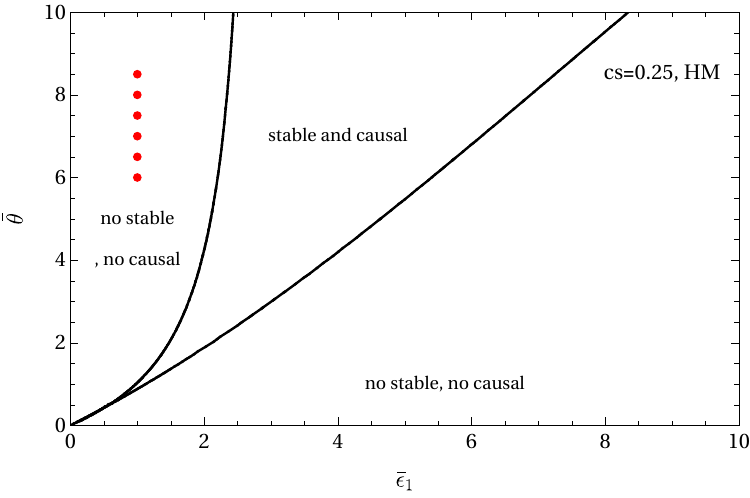}
    \\
    \vspace{0.2cm}
    \includegraphics[width=0.45\textwidth, valign=t]{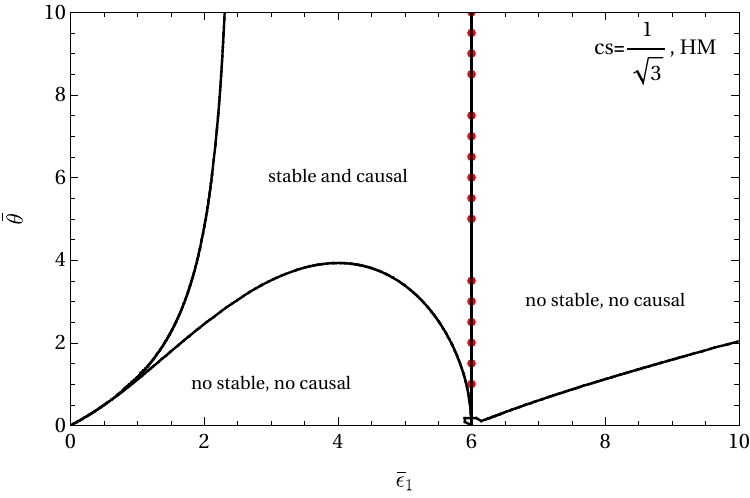}
    \hspace{0.4cm}
    \includegraphics[width=0.45\textwidth, valign=t]{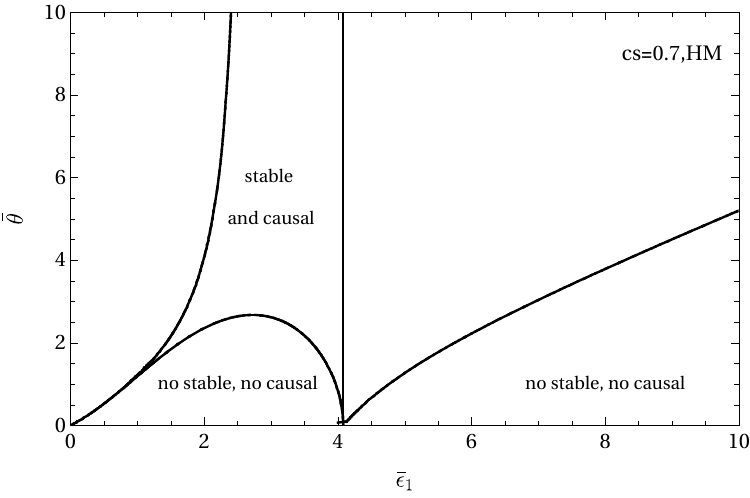}
    \caption{Comparing the stability and causality with univalence in the HM region for the BDNK model with $\bar{\epsilon}_2 = 0$ and $\bar{\pi}_1 = 3/c_s^2$. The top left (right) panels mention the $c_s = 0.1 (c_s = 0.25)$ data, and the bottom left (right) panels refer to the $c_s = 1/\sqrt{3} (c_s = 0.7)$ data. The red dots represent the points of transport space on which the Bieberbach conjecture is satisfied.}
    \label{fig: HM-BDNK}
\end{figure}

In Figure \ref{fig: LM-BDNK}, which features plots produced for $\bar{\epsilon}_2 = 0$ and $\bar{\pi}_1 = 3/c_s^2$, we illustrate the blue points in the transport space that satisfy the Bieberbach conjecture for the LM region. The top left (right) panels represent the data for $c_s = 0.1$ ($c_s = 0.25$), while the bottom left (right) panels correspond to the data for $c_s = 1/\sqrt{3}$ ($c_s = 0.7$). Each plot distinguishes between stable and causal zones as well as non-stable and non-causal zones. Notably, there are no shared points between the univalent zone and the stable and causal region in the LM part. At lower speeds, there are only a few blue points, but as the speed increases, the number of blue points also grows.
In contrast, in the IM region with $\bar{\epsilon}_2 = 0$ and $\bar{\pi}_1 = 3/c_s^2$, the scenario changes. Figure \ref{fig: IM-BDNK} displays the red points in the transport space, representing the points where the Bieberbach conjecture is satisfied in the IM region. In this case, all points within the stable and causal zone adhere to the $\bar{b}_n \leq n$ condition for $n$ ranging from 2 to 100. Unlike the LM case, the IM region features a significant number of univalent points at each $c_s$.
In Figure \ref{fig: HM-BDNK}, the red points depicted in each panel represent the points satisfying the univalent bounds in the HM region. From this data, we notice that the causal and stable region, considered the physical zone, is distinct from the univalent zone. Furthermore, the number of univalent points in the HM case is considerably fewer compared to the IM or LM situations. Additionally, the value of $c_s$ has a relatively minor impact on the number of red points observed.
\section{Relation between univalence and radius of convergence}\label{sec: rel-SC-U}
 The radius of convergence for a series representation of a spectral curve, represented by the function $F(\omega, k)$ in the complex $k$ plane, is determined by the distance between the origin and the closest point where both $F(\omega, k) = 0$ and $\partial F(\omega, k)/\partial \omega = 0$ hold simultaneously, as stated in \cite{Grozdanov:2019kge}. In this section, we aim to explore the relationship between univalence and the radius of convergence.

In the shear channel of the MIS model, as described by Eq. \eqref{eq13}, the conditions $F_\text{shear}(\omega, k) = 0$ and $\partial F_\text{shear}(\omega, k)/\partial \omega = 0$ lead to the validity of the series expansion up to the point $z_c = \frac{w}{4\eta\tau}$. On the other hand, the map in Eq. \eqref{eq: map-shear} can be solved for $\zeta$ as follows
\begin{align}
    \zeta_\pm(z) = \frac{z - 2 z_c \pm 2 \sqrt{z_c(z_c - z)}}{z}.
\end{align}
Given that $\vert \zeta \vert \leq 1$, it follows $z \leq z_c$. Consequently, the map in Eq. \eqref{eq: map-shear} is always contained within the validity region of the hydro expansion. This observation explains why univalence does not introduce any additional constraints in the shear channel. The same conclusion applies to the shear channel of the BDNK model, as described in Eq. \eqref{eq: shear-BDNK}.

However, the sound channels present a contrasting scenario, primarily due to the order of spectral equations. In the sound channel of the MIS model, as represented by the following equation (which is the dimensionless version of Eq. \eqref{eq: disp-sc-QH}):
\begin{align}\label{eq: dimless-MIS}
    F^\text{MIS}_\text{sound}(\tilde{\beta}, \tilde{z}) = \tilde{\beta}^3 + \tilde{\beta}^2 + \tilde{z} \, \tilde{\beta} (9+8X) + \tilde{z} =0,
\end{align}
and the solutions provided in Eq. \eqref{eq: sound-sol-QH}, we obtain
\begin{align}
    \frac{\partial F^\text{MIS}_\text{sound}(\tilde{\beta}, \tilde{z})}{\partial \tilde{\beta}} = 3 \tilde{\beta}^2 + 2 \tilde{\beta} + \tilde{z} (9 + 8 X) = 0,
\end{align}
where solutions take the following form
\begin{align}\label{eq: sol-betap-MIS}
   (\tilde{\beta}')_\pm = -\frac{1}{3} \left( 1 \pm \sqrt {1 - 3 \tilde{z} (9+8X)}\right).
\end{align}
Comparing Eq. \eqref{eq: sound-sol-QH} and Eq. \eqref{eq: sol-betap-MIS} reveals that at $\tilde{z} = \tilde{z}_0$ and $\tilde{z} = \tilde{z}_-$ we have $\tilde{\beta}_1(\tilde{z}) = \tilde{\beta}'_-(\tilde{z})$, while at $\tilde{z} = \tilde{z}_+$ we have $\tilde{\beta}_+(\tilde{z}) = \tilde{\beta}'_+(\tilde{z})$. As a result, $(\tilde{z}_0, \tilde{z}_\pm)$ are collision points, and the radius of convergence for a series expansion in this context is zero. This is why we obtain the sound mode dispersion relation as
\begin{align}
\omega = v k + \sum\limits_{n=2}^\infty c_n (k^2)^{\frac{n}{2}},
\end{align}
and the branch cut patterns take the form shown in Fig. \ref{fig: sing-sc-QH}. Consequently, univalence has no direct relation to the convergence in the sound channel. However, univalence serves as a new constraint to obtain a finite series in each section of the complex $\tilde{z}$ plane. For example, if $X \leq 0.853$, we can achieve a bounded series in the HM region. In this manner, we can say that hydro series is locally univalent in certain transport regions.

The situation in the sound channel of the BDNK model is quite similar. The dimensionless equation of motion is given by
\begin{align}
    F_\text{sound}^\text{BDNK} = \sum\limits_{n=0}^4 \bar{a}_n \, \bar{\beta}^n = 0,
\end{align}
where the $\bar{a}_i$ coefficients are presented in Eq. \eqref{eq: coeff-dimless-BDNK}, and the corresponding solutions can be found in Eq. \eqref{eq: on-shell-4}. From the latter equation, it can be easily demonstrated that
\begin{align}
    \frac{\partial F_\text{sound}^\text{BDNK}(\bar{\beta}, \bar{z})}{\partial \bar{\beta}} = \sum\limits_{n=0}^3 (n+1) \, \bar{a}_{n+1} \, \bar{\beta}^n = 0. 
\end{align}
The solutions to this equation are related to the branch cut patterns and the convergence properties in the sound channel of the BDNK model
\begin{align}\label{eq: betap-sol-BDNK}
 &(\bar{\beta}')_1 = - \frac{1}{4 \bar{a}_4} \bigg(\bar{a}_3 + \frac{\mathcal{K}}{\left(\mathcal{H} + \mathcal{I}\right)^{1/3}} - \left(\mathcal{H} + \mathcal{I}\right)^{1/3} \bigg),\nn\\
 & (\bar{\beta}')_s = - \frac{1}{8 \bar{a}_4} \bigg(2\bar{a}_3 - \frac{(1 + s i \sqrt{3})\mathcal{K}}{\left(\mathcal{H} + \mathcal{I}\right)^{1/3}} + (1- s i \sqrt{3})\left(\mathcal{H} + \mathcal{I}\right)^{1/3} \bigg),
\end{align}
where
\begin{align}
    & \mathcal{K} \equiv \frac{8 \bar{a}_2 \bar{a}_4}{3} - \bar{a}_3^2,\nn\\
    & \mathcal{H} \equiv - \bar{a}_3^3 + 4 \bar{a}_2 \bar{a}_3 \bar{a}_4 - 8 \bar{a}_1 \bar{a}_4^2 = \mathcal{F} \, \bar{a}^3_4, \nn\\
    & \mathcal{I}^2 \equiv  \mathcal{H}^2 + \mathcal{G}^3. 
\end{align}
At the points where $\mathcal{G}(\bar{z}) = 0$, or at the points depicted in Fig. \ref{fig: sol-BDNK}, the solutions in Eq. \eqref{eq: betap-sol-BDNK} coincide with those in Eq. \eqref{eq: on-shell-4}. This implies that the black dot points in Fig. \ref{fig: sol-BDNK} are collision points, and as a result, the sound channel of the BDNK model has a radius of convergence equal to zero. However, univalence allows for situations where, in each section of the momentum region, we can obtain a finite and bounded series representation. Due to the higher power of $\bar{\beta} = -i\omega\frac{\gamma_s}{w}$ and the increased degrees of freedom in this model compared to the MIS model, the univalence constraints can be studied on multiple variables. For instance, we can fix the momentum and examine which transports could lead to an univalent series or fix the transports and investigate the role of momentum in creating a bounded series. For our purposes, we opted for the latter approach since we aim to compare these findings with stability and causality, which are conditions that do not depend on the momentum value.
\section{Conclusion}
\label{sec5}
Stability and causality are crucial conditions in the study of high-energy dynamical systems, as they impose limitations on transport coefficients containing valuable information about the underlying microscopic theory. Relativistic hydrodynamics, an effective description of high-energy systems, is valid in both near-equilibrium and far-from-equilibrium scenarios. This macroscopic interpretation is expressed through a series of small gradients away from background values, which introduces additional mathematical constraints. While some of these constraints may not have direct physical meaning, they contribute to the self-consistency and manageability of the problem.

In this study, we explore the relationship between stability, causality, and univalence bounds to better understand the physical implications of univalence. Our analysis focuses on two quasihydrodynamic models: the MIS and the first-order hydro, also known as the BDNK model. These models are significant because they allow us to identify regions in their transport space where stability and causality are satisfied, making them suitable for implementation in hydrodynamic codes.

Our research reveals that the interplay between stability, causality, and univalence bounds depends significantly on the fluctuation propagation channel. In the shear channel, propagations occur in the transverse direction to the momentum, leading to spectral equations with frequencies raised to the power of two. We demonstrate that in this channel, the radius of convergence is non-zero, and the hydrodynamic series is univalent within this radius. Consequently, univalence offers no additional information for the shear channel of these models, as the univalence region coincides with the radius of the convergence zone.

In contrast, the sound channels present a different scenario. Higher frequency powers and specific mathematical situations, such as a zero radius of convergence, give univalence a distinct meaning. In the MIS theory with the dispersion relation provided in Eq. \eqref{eq: disp-sc-QH}, univalence imposes a constraint on $X = -1 + \gamma_s/(8 c_s^2 w \tau)$ for the high momentum region, ensuring that the hydrodynamic series maintains a bounded value despite being globally convergent. For the BDNK model, the radius of convergence is zero, and univalence does not hold globally for the series. However, due to the presence of more transports and a complex dispersion relation, univalence conditions can be examined in various ways for this model. We fix the transports and the speed of sound and analyze the role of momentum regions. Our findings reveal that in intermediate momentum regions, all stable and causal zones fit within the univalent zone. However, in low and high momentum regions, the stable and causal parts are separated from the univalent parts. This can be interpreted as follows: Locally in momentum space, we can equate stability, causality, and univalence, giving local univalence a meaningful interpretation.

This study can be extended in several directions to further explore the implications of univalence in hydrodynamic models. Firstly, it would be valuable to investigate the local univalence in other hydrodynamic models, particularly those that possess a non-vanishing space for stability and causality conditions. This would help in understanding the generality of the observed local univalence phenomenon. Secondly, examining the impact of univalence bounds on the underlying microscopic theories is an intriguing avenue for future research. Investigating the connection between univalence and many-body properties of Quantum Field Theory could provide new insights into the relationship between macroscopic and microscopic descriptions of high-energy systems. 
Lastly, understanding the role of univalence in far-from-equilibrium scenarios is crucial. By exploring the extent to which the hydrodynamic series can produce limited values in regions far away from equilibrium in both momentum and transport space, we can further refine our comprehension of the applicability and limitations of hydrodynamic descriptions in such regimes. These intriguing research directions will be pursued in our future works.
\newpage
\appendix
\section{The Coefficients \texorpdfstring{$\bar{g}_n$}{Lg}}\label{sec: App-A}
The $\bar{g}_n$ coefficients of Eq. \eqref{eq: ebar} are expressed as follows
\begin{align}
    &\bar{g}_0 = 0,\nn\\
    &\bar{g}_1 = 108 c_s^2 (\bar{\epsilon}_1-\bar{\theta} )^2, \\
    &\bar{g}_2 = 27 \bigg(20 c_s^4 \bar{\epsilon}_1^4-4 c_s^2 \bar{\epsilon}_1^3 \left(3 c_s^2 (6 \bar{\theta} +\bar{\pi}_1+\bar{\epsilon}_2)+2\right)+\bar{\epsilon}_1^2 \left(4 c_s^4 \left(19 \bar{\theta}^2+14 \bar{\theta}  \bar{\pi}_1+14 \bar{\theta}  \bar{\epsilon}_2+3 \bar{\pi}_1 \bar{\epsilon}_2\right)-16 c_s^2 \bar{\theta} -1\right)\nn\\
    &\qquad -2 \bar{\theta}  \bar{\epsilon}_1 \left(2 c_s^4 \left(8 \bar{\theta}^2+15 \bar{\theta}  \bar{\pi}_1+15 \bar{\theta}  \bar{\epsilon}_2+10 \bar{\pi}_1 \bar{\epsilon}_2\right)-24 c_s^2 \bar{\theta} -1\right)+\bar{\theta}^2 \left(4 c_s^4 \left(2 \bar{\theta}^2+4 \bar{\theta}  \bar{\pi}_1+4 \bar{\theta}  \bar{\epsilon}_2+3 \bar{\pi}_1 \bar{\epsilon}_2\right)-24 c_s^2 \bar{\theta} -1\right)\bigg), \nn\\
    &\bar{g}_3 = -54 \bigg(2 c_s^6 \bar{\epsilon}_1^6+c_s^4 \bar{\epsilon}_1^5 \left(2 c_s^2 (40 \bar{\theta} +\bar{\pi}_1+\bar{\epsilon}_2)+3\right)\nn\\
    &\qquad +c_s^4 \bar{\epsilon}_1^4 \left(-184 c_s^2 \bar{\theta}^2+\bar{\theta}  \left(51-130 c_s^2 \bar{\pi}_1\right)-6 c_s^2 \bar{\pi}_1^2-6 c_s^2 \bar{\epsilon}_2^2-2 c_s^2 \bar{\epsilon}_2 (65 \bar{\theta} +7 \bar{\pi}_1)+\bar{\pi}_1+\bar{\epsilon}_2\right)\nn\\
    & \qquad + \bar{\epsilon}_1^3 \bigg(2 c_s^6 \left(\bar{\theta}  \left(68 \bar{\theta}^2+142 \bar{\theta}  \bar{\pi}_1+27 \bar{\pi}_1^2\right)+3 \bar{\epsilon}_2^2 (9 \bar{\theta} +2 \bar{\pi}_1)+2 \bar{\epsilon}_2 \left(71 \bar{\theta}^2+53 \bar{\theta}  \bar{\pi}_1+3 \bar{\pi}_1^2\right)\right)-c_s^4 (\bar{\theta}  (175 \bar{\theta} +17 \bar{\pi}_1)+\bar{\epsilon}_2 (17 \bar{\theta} +\bar{\pi}_1))\nn\\
    &\qquad -c_s^2 (30 \bar{\theta} +\bar{\pi}_1+\bar{\epsilon}_2)-1\bigg)-\bar{\epsilon}_1^2 \bigg(52 c_s^6 \bar{\theta}^4+c_s^2 \bar{\theta}^2 \left(2 c_s^4 \left(55 \bar{\pi}_1^2+55 \bar{\epsilon}_2^2+174 \bar{\pi}_1 \bar{\epsilon}_2\right)-117 c_s^2 (\bar{\pi}_1+\bar{\epsilon}_2)-13\right)\nn\\
    &\qquad+c_s^4 \bar{\theta}^3 \left(194 c_s^2 (\bar{\pi}_1+\bar{\epsilon}_2)-141\right) +c_s^2 \bar{\pi}_1 \bar{\epsilon}_2 \left(6 c_s^4 \bar{\pi}_1 \bar{\epsilon}_2-1\right)+\bar{\theta}  \left(90 c_s^6 \bar{\pi}_1 \bar{\epsilon}_2 (\bar{\pi}_1+\bar{\epsilon}_2)-19 c_s^4 \bar{\pi}_1 \bar{\epsilon}_2-5 c_s^2 (\bar{\pi}_1+\bar{\epsilon}_2)+1\right)\bigg)\nn\\
    &\qquad +\bar{\theta}  \bar{\epsilon}_1 \bigg(12 c_s^6 \bar{\theta}^4+\bar{\theta}^3 \left(54 c_s^6 (\bar{\pi}_1+\bar{\epsilon}_2)-40 c_s^4\right)+c_s^2 \bar{\theta}^2 \left(2 c_s^4 \left(37 \bar{\pi}_1^2+37 \bar{\epsilon}_2^2+90 \bar{\pi}_1 \bar{\epsilon}_2\right)-127 c_s^2 (\bar{\pi}_1+\bar{\epsilon}_2)+30\right)\nn\\
    &\qquad + 4 c_s^2 \bar{\pi}_1 \bar{\epsilon}_2 \left(9 c_s^4 \bar{\pi}_1 \bar{\epsilon}_2-1\right)+\bar{\theta}  \left(104 c_s^6 \bar{\pi}_1 \bar{\epsilon}_2 (\bar{\pi}_1+\bar{\epsilon}_2)-91 c_s^4 \bar{\pi}_1 \bar{\epsilon}_2-5 c_s^2 (\bar{\pi}_1+\bar{\epsilon}_2)+4\right)\bigg)\nn\\
    &\qquad+\bar{\theta}^2 \bigg(-2 c_s^6 \bar{\theta}^4-c_s^2 \bar{\theta}^2 \left(2 c_s^4 \left(6 \bar{\pi}_1^2+6 \bar{\epsilon}_2^2+7 \bar{\pi}_1 \bar{\epsilon}_2\right)-26 c_s^2 (\bar{\pi}_1+\bar{\epsilon}_2)+15\right)-4 c_s^4 \bar{\theta}^3 \left(2 c_s^2 (\bar{\pi}_1+\bar{\epsilon}_2)-3\right)\nn\\
    &\qquad + c_s^2 \bar{\pi}_1 \bar{\epsilon}_2 \left(1-6 c_s^4 \bar{\pi}_1 \bar{\epsilon}_2\right)+\bar{\theta}  \left(-18 c_s^6 \bar{\pi}_1 \bar{\epsilon}_2 (\bar{\pi}_1+\bar{\epsilon}_2)+17 c_s^4 \bar{\pi}_1 \bar{\epsilon}_2+c_s^2 (\bar{\pi}_1+\bar{\epsilon}_2)-2\right)\bigg)\bigg), \nn\\
    &\bar{g}_4 = 27 \bigg(3 c_s^8 \bar{\epsilon}_1^8-2 c_s^6 \bar{\epsilon}_1^7 \left(5 c_s^2 (\bar{\epsilon}_2-2 \bar{\theta} +\bar{\pi}_1)-4\right) \nn\\
    &\qquad + c_s^4 \bar{\epsilon}_1^6 \bigg(\left(11 \bar{\epsilon}_2^2+4 (5 \bar{\theta} +8 \bar{\pi}_1) \bar{\epsilon}_2+414 \bar{\theta}^2+11 \bar{\pi}_1^2+20 \bar{\theta}  \bar{\pi}_1\right) c_s^4-2 (9 \bar{\epsilon}_2-26 \bar{\theta} +9 \bar{\pi}_1) c_s^2+6\bigg)\nn\\
    &\qquad - 2 c_s^4 \bar{\epsilon}_1^5 \bigg(2 \bar{\epsilon}_2^3 c_s^4+330 \bar{\theta}^3 c_s^4+\bar{\theta}^2 \left(544 c_s^2 \bar{\pi}_1-355\right) c_s^2+\bar{\epsilon}_2^2 \left(c_s^2 (60 \bar{\theta} +17 \bar{\pi}_1)-5\right) c_s^2+\bar{\pi}_1 \left(2 \bar{\pi}_1^2 c_s^4-5 \bar{\pi}_1 c_s^2+3\right)\nn\\
    &\qquad + 6 \bar{\theta}  \left(10 \bar{\pi}_1^2 c_s^4-3 \bar{\pi}_1 c_s^2-3\right)+\bar{\epsilon}_2 \left(\left(544 \bar{\theta}^2+128 \bar{\pi}_1 \bar{\theta} +17 \bar{\pi}_1^2\right) c_s^4-(18 \bar{\theta} +19 \bar{\pi}_1) c_s^2+3\right)\bigg) \nn\\
    &\qquad + \bar{\epsilon}_1^4 c_s^8 \bigg(\bigg(4 (20 \bar{\theta} +3 \bar{\pi}_1) \bar{\epsilon}_2^3+\left(962 \bar{\theta}^2+452 \bar{\pi}_1 \bar{\theta} +35 \bar{\pi}_1^2\right) \bar{\epsilon}_2^2+4 \left(410 \bar{\theta}^3+669 \bar{\pi}_1 \bar{\theta}^2+113 \bar{\pi}_1^2 \bar{\theta} +3 \bar{\pi}_1^3\right) \bar{\epsilon}_2\nn\\
    &\qquad + \bar{\theta}  \left(399 \bar{\theta}^3+1640 \bar{\pi}_1 \bar{\theta}^2+962 \bar{\pi}_1^2 \bar{\theta} +80 \bar{\pi}_1^3\right)\bigg) -2 c_s^6 \bigg(2 (27 \bar{\theta} +5 \bar{\pi}_1) \bar{\epsilon}_2^2+\left(537 \bar{\theta}^2+123 \bar{\pi}_1 \bar{\theta} +10 \bar{\pi}_1^2\right) \bar{\epsilon}_2\nn\\
    &\qquad + 3 \bar{\theta}  \left(209 \bar{\theta}^2+179 \bar{\pi}_1 \bar{\theta} +18 \bar{\pi}_1^2\right)\bigg) -\left(\bar{\epsilon}_2^2-4 (3 \bar{\theta} +\bar{\pi}_1) \bar{\epsilon}_2-186 \bar{\theta}^2+\bar{\pi}_1^2-12 \bar{\theta}  \bar{\pi}_1\right) c_s^4+2 (\bar{\epsilon}_2-2 \bar{\theta} +\bar{\pi}_1) c_s^2-1\bigg) \nn\\
    &\qquad - 2 \bar{\epsilon}_1^3 c_s^4 \bigg(96 \bar{\theta}^5 c_s^8 + \bar{\theta}^4 \left(415 c_s^2 (\bar{\epsilon}_2+\bar{\pi}_1)-297\right) c_s^6+\bar{\theta}^3 \left(2 \left(337 \bar{\epsilon}_2^2+772 \bar{\pi}_1 \bar{\epsilon}_2+337 \bar{\pi}_1^2\right) c_s^4-913 (\bar{\epsilon}_2+\bar{\pi}_1) c_s^2+264\right) \nn\\
    &\qquad + \bar{\epsilon}_2 \bar{\pi}_1 c_s^2 \left(6 \bar{\epsilon}_2 \bar{\pi}_1 (\bar{\epsilon}_2+\bar{\pi}_1) c_s^6-5 \bar{\epsilon}_2 \bar{\pi}_1 c_s^4-(\bar{\epsilon}_2+\bar{\pi}_1) c_s^2+1\right) \nn\\
    &\qquad  + \bar{\theta}^2 c_s^2\left(2 \left(70 \bar{\epsilon}_2^3+537 \bar{\pi}_1 \bar{\epsilon}_2^2+537 \bar{\pi}_1^2 \bar{\epsilon}_2+70 \bar{\pi}_1^3\right) c_s^6-2 \left(100 \bar{\epsilon}_2^2+391 \bar{\pi}_1 \bar{\epsilon}_2+100 \bar{\pi}_1^2\right) c_s^4-18 (\bar{\epsilon}_2+\bar{\pi}_1) c_s^2+51\right) \nn\\
    &\qquad+\bar{\theta}  \left(2 \bar{\epsilon}_2 \bar{\pi}_1 \left(54 \bar{\epsilon}_2^2+131 \bar{\pi}_1 \bar{\epsilon}_2+54 \bar{\pi}_1^2\right) c_s^8-95 \bar{\epsilon}_2 \bar{\pi}_1 (\bar{\epsilon}_2+\bar{\pi}_1) c_s^6-6 \left(\bar{\epsilon}_2^2+\bar{\pi}_1 \bar{\epsilon}_2+\bar{\pi}_1^2\right) c_s^4+2 (\bar{\epsilon}_2+\bar{\pi}_1) c_s^2+4\right)\bigg) \nn\\
    &\qquad + \bar{\epsilon}_1^2 \bigg(48 \bar{\theta}^6 c_s^8+6 \bar{\theta}^5 \left(50 c_s^2 (\bar{\epsilon}_2+\bar{\pi}_1)-37\right) c_s^6+\bar{\epsilon}_2^2 \bar{\pi}_1^2 \left(4 c_s^4 \bar{\epsilon}_2 \bar{\pi}_1-1\right) c_s^4\nn
    \end{align}
    \begin{align}
    &\qquad + \bar{\theta}^4 c_s^4 \left(\left(611 \bar{\epsilon}_2^2+1204 \bar{\pi}_1 \bar{\epsilon}_2+611 \bar{\pi}_1^2\right) c_s^4-832 (\bar{\epsilon}_2+\bar{\pi}_1) c_s^2+207\right)\nn\\
    &\qquad + 2 \bar{\epsilon}_2 \bar{\theta}  \bar{\pi}_1 c_s^2\left(96 \bar{\epsilon}_2 \bar{\pi}_1 (\bar{\epsilon}_2+\bar{\pi}_1) c_s^6-41 \bar{\epsilon}_2 \bar{\pi}_1 c_s^4-11 (\bar{\epsilon}_2+\bar{\pi}_1) c_s^2+1\right) \nn\\
    &\qquad+2 \bar{\theta}^3 c_s^2 \left(16 \left(11 \bar{\epsilon}_2^3+58 \bar{\pi}_1 \bar{\epsilon}_2^2+58 \bar{\pi}_1^2 \bar{\epsilon}_2+11 \bar{\pi}_1^3\right) c_s^6-2 \left(157 \bar{\epsilon}_2^2+459 \bar{\pi}_1 \bar{\epsilon}_2+157 \bar{\pi}_1^2\right) c_s^4+83 (\bar{\epsilon}_2+\bar{\pi}_1) c_s^2+33\right) \nn\\
    &\qquad + 2 \bar{\theta}^2 \left(3 \bar{\epsilon}_2 \bar{\pi}_1 \left(88 \bar{\epsilon}_2^2+241 \bar{\pi}_1 \bar{\epsilon}_2+88 \bar{\pi}_1^2\right) c_s^8-283 \bar{\epsilon}_2 \bar{\pi}_1 (\bar{\epsilon}_2+\bar{\pi}_1) c_s^6-\left(11 \bar{\epsilon}_2^2+26 \bar{\pi}_1 \bar{\epsilon}_2+11 \bar{\pi}_1^2\right) c_s^4+11 (\bar{\epsilon}_2+\bar{\pi}_1) c_s^2+4\right)\bigg)\nn\\
    &\qquad -2 c_s^2 \bar{\theta}  \bar{\epsilon}_1 \bigg(2 \bar{\epsilon}_2^3 c_s^6 \left(41 \bar{\theta}^3+90 \bar{\pi}_1 \bar{\theta}^2+63 \bar{\pi}_1^2 \bar{\theta} +14 \bar{\pi}_1^3\right) \nn\\
    &\qquad + \bar{\epsilon}_2^2 c_s^2 \left(70 c_s^4 \bar{\theta}^4+c_s^2 \left(269 c_s^2 \bar{\pi}_1-179\right) \bar{\theta}^3+\left(322 \bar{\pi}_1^2 c_s^4-261 \bar{\pi}_1 c_s^2-6\right) \bar{\theta}^2+\bar{\pi}_1 \left(126 \bar{\pi}_1^2 c_s^4-91 \bar{\pi}_1 c_s^2-11\right) \bar{\theta} -5 \bar{\pi}_1^2\right)\nn\\
    &\qquad + \bar{\theta}^2 \left(24 \bar{\theta}^3 \left(c_s^2 \bar{\pi}_1-1\right) c_s^4+5 \bar{\theta}^2 \left(14 \bar{\pi}_1^2 c_s^4-21 \bar{\pi}_1 c_s^2+3\right) c_s^2+2 \bar{\pi}_1 \left(5-3 c_s^2 \bar{\pi}_1\right)+\bar{\theta}  \left(82 \bar{\pi}_1^3 c_s^6-179 \bar{\pi}_1^2 c_s^4+109 \bar{\pi}_1 c_s^2-6\right)\right)\nn\\
    &\qquad + \bar{\epsilon}_2 \bar{\theta}  \bigg(24 \bar{\theta}^4 c_s^6+35 \bar{\theta}^3 \left(4 c_s^2 \bar{\pi}_1-3\right) c_s^4+\bar{\theta}^2 \left(269 \bar{\pi}_1^2 c_s^4-439 \bar{\pi}_1 c_s^2+109\right) c_s^2+\bar{\pi}_1 \left(10-11 c_s^2 \bar{\pi}_1\right)\nn\\
    &\qquad + \bar{\theta}  \left(180 \bar{\pi}_1^3 c_s^6-261 \bar{\pi}_1^2 c_s^4+82 \bar{\pi}_1 c_s^2+10\right)\bigg)\bigg) + c_s^4 \bar{\theta}^2 \bigg(4 \bar{\epsilon}_2^3 (\bar{\theta} +\bar{\pi}_1)^2 (4 \bar{\theta} +\bar{\pi}_1) c_s^4\nn\\
    &\qquad + \bar{\theta}^2 \bigg(\left(8 \bar{\pi}_1^2 c_s^4-36 \bar{\pi}_1 c_s^2+27\right) \bar{\theta}^2+2 \bar{\pi}_1 \left(8 \bar{\pi}_1^2 c_s^4-16 \bar{\pi}_1 c_s^2+9\right) \bar{\theta} -\bar{\pi}_1^2\bigg)\nn\\
    &\qquad + 2 \bar{\epsilon}_2 \bar{\theta}  \bigg(2 c_s^2 \left(4 c_s^2 \bar{\pi}_1-9\right) \bar{\theta}^3+\left(14 \bar{\pi}_1^2 c_s^4-23 \bar{\pi}_1 c_s^2+9\right) \bar{\theta}^2+\bar{\pi}_1 \left(18 \bar{\pi}_1^2 c_s^4-23 \bar{\pi}_1 c_s^2+8\right) \bar{\theta} -\bar{\pi}_1^2\bigg)\nn\\
    &\qquad + \bar{\epsilon}_2^2 \left(8 c_s^4 \bar{\theta}^4+4 c_s^2 \left(7 c_s^2 \bar{\pi}_1-8\right) \bar{\theta}^3+\left(71 \bar{\pi}_1^2 c_s^4-46 \bar{\pi}_1 c_s^2-1\right) \bar{\theta}^2+2 \bar{\pi}_1 \left(12 \bar{\pi}_1^2 c_s^4-7 \bar{\pi}_1 c_s^2-1\right) \bar{\theta} -\bar{\pi}_1^2\right)\bigg)\bigg), \nn\\
    &\bar{g}_5 = -108 \bar{\theta}  \bigg(4 \bar{\epsilon}_1^9 c_s^{10} - \bar{\theta}^2 c_s^8\left(c_s^2 (\bar{\epsilon}_2+\bar{\pi}_1)-1\right) (\bar{\theta}  \bar{\pi}_1+\bar{\epsilon}_2 (\bar{\theta} +\bar{\pi}_1))^3 \nn\\
    &\quad + \bar{\epsilon}_1^8 c_s^8 \left((-17 \bar{\epsilon}_2+4 \bar{\theta} -17 \bar{\pi}_1) c_s^2+15\right) +\bar{\epsilon}_1^7 c_s^6\left(\left(27 \bar{\epsilon}_2^2+18 \bar{\theta}  \bar{\epsilon}_2+70 \bar{\pi}_1 \bar{\epsilon}_2+104 \bar{\theta}^2+27 \bar{\pi}_1^2+18 \bar{\theta}  \bar{\pi}_1\right) c_s^4-16 (3 \bar{\epsilon}_2-\bar{\theta} +3 \bar{\pi}_1) c_s^2+20\right)\nn\\
    &\qquad + \bar{\epsilon}_1 c_s^6 \left(\bar{\theta}  \bar{\pi}_1+\bar{\epsilon}_2 (\bar{\theta} +\bar{\pi}_1)\right)  \times  \bigg(\bar{\epsilon}_2^3 \left(21 \bar{\theta}^3+43 \bar{\pi}_1 \bar{\theta}^2+26 \bar{\pi}_1^2 \bar{\theta} +4 \bar{\pi}_1^3\right) c_s^4+\bar{\theta}^2 \bigg(4 \left(4 \bar{\pi}_1^2 c_s^4-13 \bar{\pi}_1 c_s^2+9\right) \bar{\theta}^2\nn\\
    &\qquad + \bar{\pi}_1 \bar{\theta} \left(21 \bar{\pi}_1^2 c_s^4-42 \bar{\pi}_1 c_s^2+23\right)  -\bar{\pi}_1^2\bigg)+\bar{\epsilon}_2 \bar{\theta}  \bigg(4 c_s^2 \left(8 c_s^2 \bar{\pi}_1-13\right) \bar{\theta}^3+\left(43 \bar{\pi}_1^2 c_s^4-64 \bar{\pi}_1 c_s^2+23\right) \bar{\theta}^2\nn\\
    &\qquad + \bar{\pi}_1 \left(43 \bar{\pi}_1^2 c_s^4-58 \bar{\pi}_1 c_s^2+21\right) \bar{\theta} -2 \bar{\pi}_1^2\bigg)+\bar{\epsilon}_2^2 \bigg(16 c_s^4 \bar{\theta}^4+c_s^2 \left(43 c_s^2 \bar{\pi}_1-42\right) \bar{\theta}^3\nn\\
    &\qquad + \left(86 \bar{\pi}_1^2 c_s^4-58 \bar{\pi}_1 c_s^2-1\right) \bar{\theta}^2+2 \bar{\pi}_1 \left(13 \bar{\pi}_1^2 c_s^4-8 \bar{\pi}_1 c_s^2-1\right) \bar{\theta} -\bar{\pi}_1^2\bigg)\bigg) \nn\\
    &\qquad - \bar{\epsilon}_1^6 c_s^4 \bigg(\left(19 \bar{\epsilon}_2^3+27 (3 \bar{\theta} +4 \bar{\pi}_1) \bar{\epsilon}_2^2+\left(357 \bar{\theta}^2+170 \bar{\pi}_1 \bar{\theta} +108 \bar{\pi}_1^2\right) \bar{\epsilon}_2+92 \bar{\theta}^3+19 \bar{\pi}_1^3+81 \bar{\theta}  \bar{\pi}_1^2+357 \bar{\theta}^2 \bar{\pi}_1\right) c_s^6\nn\\
    &\qquad - 3 c_s^4 \left(17 \bar{\epsilon}_2^2+17 \bar{\theta}  \bar{\epsilon}_2+49 \bar{\pi}_1 \bar{\epsilon}_2+99 \bar{\theta}^2+17 \bar{\pi}_1^2+17 \bar{\theta}  \bar{\pi}_1\right) +6 (7 \bar{\epsilon}_2-4 \bar{\theta} +7 \bar{\pi}_1) c_s^2-10\bigg)\nn\\
    &\qquad +\bar{\epsilon}_1^5 c_s^4 \bigg(5 \bar{\epsilon}_2^4 c_s^6+96 \bar{\theta}^4 c_s^6+4 \bar{\theta}^3 \left(91 c_s^2 \bar{\pi}_1-69\right) c_s^4+2 \bar{\epsilon}_2^3 \left(c_s^2 (46 \bar{\theta} +37 \bar{\pi}_1)-9\right) c_s^4+\bar{\theta}^2 c_s^2 \left(469 \bar{\pi}_1^2 c_s^4-714 \bar{\pi}_1 c_s^2+267\right) \nn\\
    &\qquad + \bar{\epsilon}_2^2 \left(\left(469 \bar{\theta}^2+382 \bar{\pi}_1 \bar{\theta} +162 \bar{\pi}_1^2\right) c_s^4-6 (26 \bar{\theta} +25 \bar{\pi}_1) c_s^2+21\right) c_s^2+\bar{\pi}_1 \left(c_s^2 \bar{\pi}_1-1\right)^2 \left(5 c_s^2 \bar{\pi}_1-8\right)\nn\\
    &\qquad + \bar{\theta}  \left(92 \bar{\pi}_1^3 c_s^6-156 \bar{\pi}_1^2 c_s^4+45 \bar{\pi}_1 c_s^2+16\right)\nn\\
    &\qquad + \bar{\epsilon}_2 \left(2 \left(182 \bar{\theta}^3+573 \bar{\pi}_1 \bar{\theta}^2+191 \bar{\pi}_1^2 \bar{\theta} +37 \bar{\pi}_1^3\right) c_s^6-6 \left(119 \bar{\theta}^2+56 \bar{\pi}_1 \bar{\theta} +25 \bar{\pi}_1^2\right) c_s^4+9 (5 \bar{\theta} +9 \bar{\pi}_1) c_s^2-8\right)\bigg)\nn\\
    &\qquad - \bar{\epsilon}_1^2 c_s^4 \bigg(\bar{\epsilon}_2^4 c_s^6\left(62 \bar{\theta}^3+138 \bar{\pi}_1 \bar{\theta}^2+93 \bar{\pi}_1^2 \bar{\theta} +17 \bar{\pi}_1^3\right) + \bar{\epsilon}_2^3 c_s^2 \bigg(111 c_s^4 \bar{\theta}^4+2 c_s^2 \left(231 c_s^2 \bar{\pi}_1-67\right) \bar{\theta}^3+\left(569 \bar{\pi}_1^2 c_s^4-234 \bar{\pi}_1 c_s^2-3\right) \bar{\theta}^2 \nn\\
    &\qquad +\bar{\pi}_1 \left(202 \bar{\pi}_1^2 c_s^4-115 \bar{\pi}_1 c_s^2-6\right) \bar{\theta} +\bar{\pi}_1^2 \left(17 \bar{\pi}_1^2 c_s^4-15 \bar{\pi}_1 c_s^2-3\right)\bigg)\nn\\
    &\qquad + \bar{\theta}^2 \bigg(4 c_s^2 \left(16 \bar{\pi}_1^2 c_s^4-33 \bar{\pi}_1 c_s^2+12\right) \bar{\theta}^3+\left(111 \bar{\pi}_1^3 c_s^6-163 \bar{\pi}_1^2 c_s^4+36 \bar{\pi}_1 c_s^2+12\right) \bar{\theta}^2+\bar{\pi}_1 \left(62 \bar{\pi}_1^3 c_s^6-134 \bar{\pi}_1^2 c_s^4+81 \bar{\pi}_1 c_s^2-7\right) \bar{\theta}\nn\\
    &\qquad - 3 \bar{\pi}_1^2 \left(c_s^2 \bar{\pi}_1-1\right)\bigg)+\bar{\epsilon}_2 \bar{\theta}  \bigg(4 c_s^4 \left(32 c_s^2 \bar{\pi}_1-33\right) \bar{\theta}^4+c_s^2 \left(329 \bar{\pi}_1^2 c_s^4-362 \bar{\pi}_1 c_s^2+36\right) \bar{\theta}^3+\left(462 \bar{\pi}_1^3 c_s^6-632 \bar{\pi}_1^2 c_s^4+186 \bar{\pi}_1 c_s^2-7\right) \bar{\theta}^2\nn\\
    &\qquad + \bar{\pi}_1 \left(138 \bar{\pi}_1^3 c_s^6-234 \bar{\pi}_1^2 c_s^4+100 \bar{\pi}_1 c_s^2-1\right) \bar{\theta} -6 \bar{\pi}_1^2 \left(c_s^2 \bar{\pi}_1-1\right)\bigg)\nn
    \end{align}
    \begin{align}
    &\qquad + \bar{\epsilon}_2^2 \bigg(64 \bar{\theta}^5 c_s^6+\bar{\theta}^4 \left(329 c_s^2 \bar{\pi}_1-163\right) c_s^4+\bar{\theta}^3 \left(780 \bar{\pi}_1^2 c_s^4-632 \bar{\pi}_1 c_s^2+81\right) c_s^2-3 \bar{\pi}_1^2 \left(c_s^2 \bar{\pi}_1-1\right)\nn\\
    &\qquad+\bar{\theta}  \bar{\pi}_1 \left(93 \bar{\pi}_1^3 c_s^6-115 \bar{\pi}_1^2 c_s^4+16 \bar{\pi}_1 c_s^2+6\right)+\bar{\theta}^2 \left(569 \bar{\pi}_1^3 c_s^6-595 \bar{\pi}_1^2 c_s^4+100 \bar{\pi}_1 c_s^2+3\right)\bigg)\bigg) \nn\\
    &\qquad + \bar{\epsilon}_1^3 c_s^2 \bigg(\bar{\epsilon}_2^4 \left(70 \bar{\theta}^2+96 \bar{\pi}_1 \bar{\theta} +27 \bar{\pi}_1^2\right) c_s^8\nn\\
    &\qquad +\bar{\epsilon}_2^3 c_s^4 \left(308 \bar{\theta}^3 c_s^4+2 \bar{\theta}^2 \left(359 c_s^2 \bar{\pi}_1-82\right) c_s^2+\bar{\pi}_1 \left(70 \bar{\pi}_1^2 c_s^4-48 \bar{\pi}_1 c_s^2-3\right)+\bar{\theta}  \left(406 \bar{\pi}_1^2 c_s^4-188 \bar{\pi}_1 c_s^2-3\right)\right)\nn\\
    &\qquad + \bar{\epsilon}_2^2 c_s^2 \bigg(248 \bar{\theta}^4 c_s^6+4 \bar{\theta}^3 \left(277 c_s^2 \bar{\pi}_1-150\right) c_s^4+2 \bar{\theta}^2 \left(700 \bar{\pi}_1^2 c_s^4-556 \bar{\pi}_1 c_s^2+53\right) c_s^2+3 \bar{\pi}_1 \left(9 \bar{\pi}_1^3 c_s^6-16 \bar{\pi}_1^2 c_s^4+4 \bar{\pi}_1 c_s^2+2\right)\nn\\
    &\qquad + \bar{\theta}  \left(406 \bar{\pi}_1^3 c_s^6-432 \bar{\pi}_1^2 c_s^4+85 \bar{\pi}_1 c_s^2+6\right)\bigg) +\bar{\theta}  \bigg(96 \bar{\theta}^4 \left(c_s^2 \bar{\pi}_1-1\right) c_s^6+8 \bar{\theta}^3 \left(31 \bar{\pi}_1^2 c_s^4-36 \bar{\pi}_1 c_s^2+9\right) c_s^4-3 \bar{\pi}_1 \left(c_s^2 \bar{\pi}_1-1\right)^2\nn\\
    &\qquad + \bar{\theta}^2 \left(308 \bar{\pi}_1^3 c_s^8-600 \bar{\pi}_1^2 c_s^6+378 \bar{\pi}_1 c_s^4-92 c_s^2\right)+\bar{\theta}  \left(70 \bar{\pi}_1^4 c_s^8-164 \bar{\pi}_1^3 c_s^6+106 \bar{\pi}_1^2 c_s^4-15\right)\bigg)\nn\\
    &\qquad +\bar{\epsilon}_2 \bigg(96 \bar{\theta}^5 c_s^8+16 \bar{\theta}^4 \left(31 c_s^2 \bar{\pi}_1-18\right) c_s^6+2 \bar{\theta}^2 \bar{\pi}_1 \left(359 \bar{\pi}_1^2 c_s^4-556 \bar{\pi}_1 c_s^2+217\right) c_s^4+2 \bar{\theta}^3 \left(554 \bar{\pi}_1^2 c_s^4-624 \bar{\pi}_1 c_s^2+189\right) c_s^4\nn\\
    &\qquad - 3 \bar{\pi}_1 \left(c_s^2 \bar{\pi}_1-1\right)^2+\bar{\theta}  \left(96 \bar{\pi}_1^4 c_s^8-188 \bar{\pi}_1^3 c_s^6+85 \bar{\pi}_1^2 c_s^4+4 \bar{\pi}_1 c_s^2-3\right)\bigg)\bigg)\nn\\
    &\qquad - \bar{\epsilon}_1^4 \bigg(c_s^{10} \bigg((33 \bar{\theta} +19 \bar{\pi}_1) \bar{\epsilon}_2^4+2 \left(143 \bar{\theta}^2+163 \bar{\pi}_1 \bar{\theta} +54 \bar{\pi}_1^2\right) \bar{\epsilon}_2^3+\left(519 \bar{\theta}^3+1369 \bar{\pi}_1 \bar{\theta}^2+614 \bar{\pi}_1^2 \bar{\theta} +108 \bar{\pi}_1^3\right) \bar{\epsilon}_2^2\nn\\
    &\qquad + \left(252 \bar{\theta}^4+1062 \bar{\pi}_1 \bar{\theta}^3+1369 \bar{\pi}_1^2 \bar{\theta}^2+326 \bar{\pi}_1^3 \bar{\theta} +19 \bar{\pi}_1^4\right) \bar{\epsilon}_2+\bar{\theta}  \left(48 \bar{\theta}^4+252 \bar{\pi}_1 \bar{\theta}^3+519 \bar{\pi}_1^2 \bar{\theta}^2+286 \bar{\pi}_1^3 \bar{\theta} +33 \bar{\pi}_1^4\right)\bigg)\nn\\
    &\qquad - c_s^8 \bigg((89 \bar{\theta} +51 \bar{\pi}_1) \bar{\epsilon}_2^3+\left(578 \bar{\theta}^2+473 \bar{\pi}_1 \bar{\theta} +147 \bar{\pi}_1^2\right) \bar{\epsilon}_2^2+\left(735 \bar{\theta}^3+1579 \bar{\pi}_1 \bar{\theta}^2+473 \bar{\pi}_1^2 \bar{\theta} +51 \bar{\pi}_1^3\right) \bar{\epsilon}_2\nn\\
    &\qquad + \bar{\theta}  \left(180 \bar{\theta}^3+735 \bar{\pi}_1 \bar{\theta}^2+578 \bar{\pi}_1^2 \bar{\theta} +89 \bar{\pi}_1^3\right)\bigg) \nn\\
    &\qquad + c_s^6 \bigg(-\bar{\epsilon}_2^3+(69 \bar{\theta} +36 \bar{\pi}_1) \bar{\epsilon}_2^2+3 \left(119 \bar{\theta}^2+54 \bar{\pi}_1 \bar{\theta} +12 \bar{\pi}_1^2\right) \bar{\epsilon}_2+276 \bar{\theta}^3-\bar{\pi}_1^3+69 \bar{\theta}  \bar{\pi}_1^2+357 \bar{\theta}^2 \bar{\pi}_1\bigg)\nn\\
    &\qquad +\left(3 \bar{\epsilon}_2^2-(9 \bar{\theta} +\bar{\pi}_1) \bar{\epsilon}_2-59 \bar{\theta}^2+3 \bar{\pi}_1^2-9 \bar{\theta}  \bar{\pi}_1\right) c_s^4-(3 \bar{\epsilon}_2+4 \bar{\theta} +3 \bar{\pi}_1) c_s^2+1\bigg)\bigg), \nn\\
    &\bar{g}_6 = 432 c_s^2 \bar{\theta}^2 \bar{\epsilon}_1 \left(c_s^2 (-\bar{\pi}_1+\bar{\epsilon}_1-\bar{\epsilon}_2)+1\right) \nn\\
    &\quad \times \bigg(c_s^4 \bar{\epsilon}_1^4+c_s^4 \left(\bar{\theta}  \bar{\pi}_1+\bar{\epsilon}_2 (\bar{\theta} +\bar{\pi}_1)\right)^2 - 2 c_s^2 \bar{\epsilon}_1^3 \left(c_s^2 (\bar{\pi}_1+\bar{\epsilon}_2)-1\right)-2 c_s^2 \bar{\epsilon}_1 \left(c_s^2 (\bar{\pi}_1+\bar{\epsilon}_2)-1\right) (\bar{\theta}  (2 \bar{\theta} +\bar{\pi}_1)+ \bar{\epsilon}_2 (\bar{\theta} +\bar{\pi}_1))\nn\\
    &\qquad +\bar{\epsilon}_1^2 \bigg(c_s^4 \left(4 \bar{\theta}^2+2 \bar{\theta}  \bar{\pi}_1+\bar{\pi}_1^2+\bar{\epsilon}_2^2+2 \bar{\epsilon}_2 (\bar{\theta} +2 \bar{\pi}_1)\right)-2 c_s^2 (\bar{\pi}_1+\bar{\epsilon}_2)+1\bigg)\bigg)^2.\nn
\end{align}
\bibliographystyle{fullsort}
\bibliography{refs}

\providecommand{\href}[2]{#2}\begingroup\raggedright\begin{thebibliography}{10}

\bibitem{Rischke:2003mt}
D.~H. Rischke, ``{The Quark gluon plasma in equilibrium},'' {\em Prog. Part. Nucl. Phys.} {\bf 52} (2004) 197--296, \href{http://www.arXiv.org/abs/nucl-th/0305030}{{\tt nucl-th/0305030}}.

\bibitem{Annala:2019puf}
E.~Annala, T.~Gorda, A.~Kurkela, J.~N\"attil\"a, and A.~Vuorinen, ``{Evidence for quark-matter cores in massive neutron stars},'' {\em Nature Phys.} {\bf 16} (2020), no.~9, 907--910, \href{http://www.arXiv.org/abs/1903.09121}{{\tt 1903.09121}}.

\bibitem{Shuryak:2004cy}
E.~V. Shuryak, ``{What RHIC experiments and theory tell us about properties of quark-gluon plasma?},'' {\em Nucl. Phys. A} {\bf 750} (2005) 64--83, \href{http://www.arXiv.org/abs/hep-ph/0405066}{{\tt hep-ph/0405066}}.

\bibitem{Shuryak:2003xe}
E.~Shuryak, ``{Why does the quark gluon plasma at RHIC behave as a nearly ideal fluid?},'' {\em Prog. Part. Nucl. Phys.} {\bf 53} (2004) 273--303, \href{http://www.arXiv.org/abs/hep-ph/0312227}{{\tt hep-ph/0312227}}.

\bibitem{Busza:2018rrf}
W.~Busza, K.~Rajagopal, and W.~van~der Schee, ``{Heavy Ion Collisions: The Big Picture, and the Big Questions},'' {\em Ann. Rev. Nucl. Part. Sci.} {\bf 68} (2018) 339--376, \href{http://www.arXiv.org/abs/1802.04801}{{\tt 1802.04801}}.

\bibitem{Heinz:2013th}
U.~Heinz and R.~Snellings, ``{Collective flow and viscosity in relativistic heavy-ion collisions},'' {\em Ann. Rev. Nucl. Part. Sci.} {\bf 63} (2013) 123--151, \href{http://www.arXiv.org/abs/1301.2826}{{\tt 1301.2826}}.

\bibitem{Teaney:2000cw}
D.~Teaney, J.~Lauret, and E.~V. Shuryak, ``{Flow at the SPS and RHIC as a quark gluon plasma signature},'' {\em Phys. Rev. Lett.} {\bf 86} (2001) 4783--4786, \href{http://www.arXiv.org/abs/nucl-th/0011058}{{\tt nucl-th/0011058}}.

\bibitem{Heller:2015dha}
M.~P. Heller and M.~Spali{\'n}ski, ``{Hydrodynamics Beyond the Gradient Expansion: Resurgence and Resummation},'' {\em Phys. Rev. Lett.} {\bf 115} (2015), no.~7, 072501, \href{http://www.arXiv.org/abs/1503.07514}{{\tt 1503.07514}}.

\bibitem{Heller:2016rtz}
M.~P. Heller, A.~Kurkela, M.~Spaliński, and V.~Svensson, ``{Hydrodynamization in kinetic theory: Transient modes and the gradient expansion},'' {\em Phys. Rev.} {\bf D97} (2018), no.~9, 091503,
\href{http://www.arXiv.org/abs/1609.04803}{{\tt 1609.04803}}.

\bibitem{Heller:2018qvh}
M.~P. Heller and V.~Svensson, ``{How does relativistic kinetic theory remember about initial conditions?},'' {\em Phys. Rev. D} {\bf 98} (2018), no.~5, 054016, \href{http://www.arXiv.org/abs/1802.08225}{{\tt 1802.08225}}.

\bibitem{Romatschke:2017vte}
P.~Romatschke, ``{Relativistic Fluid Dynamics Far From Local Equilibrium},'' {\em Phys. Rev. Lett.} {\bf 120} (2018), no.~1, 012301, \href{http://www.arXiv.org/abs/1704.08699}{{\tt 1704.08699}}.

\bibitem{CMS:2015yux}
{\bf CMS} Collaboration, V.~Khachatryan {\em et al.}, ``{Evidence for Collective Multiparticle Correlations in p-Pb Collisions},'' {\em Phys. Rev. Lett.} {\bf 115} (2015), no.~1, 012301, \href{http://www.arXiv.org/abs/1502.05382}{{\tt 1502.05382}}.

\bibitem{CMS:2016fnw}
{\bf CMS} Collaboration, V.~Khachatryan {\em et al.}, ``{Evidence for collectivity in pp collisions at the LHC},'' {\em Phys. Lett. B} {\bf 765} (2017) 193--220, \href{http://www.arXiv.org/abs/1606.06198}{{\tt 1606.06198}}.

\bibitem{ATLAS:2015hzw}
{\bf ATLAS} Collaboration, G.~Aad {\em et al.}, ``{Observation of Long-Range Elliptic Azimuthal Anisotropies in $\sqrt{s}=$13 and 2.76 TeV $pp$ Collisions with the ATLAS Detector},'' {\em Phys. Rev. Lett.} {\bf 116} (2016), no.~17, 172301, \href{http://www.arXiv.org/abs/1509.04776}{{\tt 1509.04776}}.

\bibitem{Noronha-Hostler:2015wft}
J.~Noronha-Hostler, J.~Noronha, and M.~Gyulassy, ``{The unreasonable effectiveness of hydrodynamics in heavy ion collisions},'' {\em Nucl. Phys. A} {\bf 956} (2016) 890--893, \href{http://www.arXiv.org/abs/1512.07135}{{\tt 1512.07135}}.

\bibitem{Hiscock:1983zz}
W.~A. Hiscock and L.~Lindblom, ``{Stability and causality in dissipative relativistic fluids},'' {\em Annals Phys.} {\bf 151} (1983) 466--496.

\bibitem{Pu:2009fj}
S.~Pu, T.~Koide, and D.~H. Rischke, ``{Does stability of relativistic dissipative fluid dynamics imply causality?},'' {\em Phys. Rev. D} {\bf 81} (2010) 114039, \href{http://www.arXiv.org/abs/0907.3906}{{\tt 0907.3906}}.

\bibitem{Gavassino:2023mad}
L.~Gavassino, M.~M. Disconzi, and J.~Noronha, ``{Dispersion Relations Alone Cannot Guarantee Causality},'' {\em Phys. Rev. Lett.} {\bf 132} (2024), no.~16, 162301, \href{http://www.arXiv.org/abs/2307.05987}{{\tt 2307.05987}}.

\bibitem{Wang:2023csj}
D.-L. Wang and S.~Pu, ``{Stability and causality criteria in linear mode analysis: Stability means causality},'' {\em Phys. Rev. D} {\bf 109} (2024), no.~3, L031504, \href{http://www.arXiv.org/abs/2309.11708}{{\tt 2309.11708}}.

\bibitem{Hoult:2023clg}
R.~E. Hoult and P.~Kovtun, ``{Causality and classical dispersion relations},'' {\em Phys. Rev. D} {\bf 109} (2024), no.~4, 046018, \href{http://www.arXiv.org/abs/2309.11703}{{\tt 2309.11703}}.

\bibitem{Mullins:2023ott}
N.~Mullins, M.~Hippert, L.~Gavassino, and J.~Noronha, ``{Relativistic hydrodynamic fluctuations from an effective action: Causality, stability, and the information current},'' {\em Phys. Rev. D} {\bf 108} (2023), no.~11, 116019, \href{http://www.arXiv.org/abs/2309.00512}{{\tt 2309.00512}}.

\bibitem{Gavassino:2023odx}
L.~Gavassino, M.~M. Disconzi, and J.~Noronha, ``{Universality Classes of Relativistic Fluid Dynamics I: Foundations},'' \href{http://www.arXiv.org/abs/2302.03478}{{\tt 2302.03478}}.

\bibitem{Gavassino:2023qwl}
L.~Gavassino, M.~M. Disconzi, and J.~Noronha, ``{Universality Classes of Relativistic Fluid Dynamics II: Applications},'' \href{http://www.arXiv.org/abs/2302.05332}{{\tt 2302.05332}}.

\bibitem{Grozdanov:2019kge}
S.~Grozdanov, P.~K. Kovtun, A.~O. Starinets, and P.~Tadi\'c, ``{Convergence of the Gradient Expansion in Hydrodynamics},'' {\em Phys. Rev. Lett.} {\bf 122} (2019), no.~25, 251601, \href{http://www.arXiv.org/abs/1904.01018}{{\tt 1904.01018}}.

\bibitem{Withers:2018srf}
B.~Withers, ``{Short-lived modes from hydrodynamic dispersion relations},'' {\em JHEP} {\bf 06} (2018) 059, \href{http://www.arXiv.org/abs/1803.08058}{{\tt 1803.08058}}.

\bibitem{Abbasi:2020ykq}
N.~Abbasi and S.~Tahery, ``{Complexified quasinormal modes and the pole-skipping in a holographic system at finite chemical potential},'' {\em JHEP} {\bf 10} (2020) 076, \href{http://www.arXiv.org/abs/2007.10024}{{\tt 2007.10024}}.

\bibitem{Jansen:2020hfd}
A.~Jansen and C.~Pantelidou, ``{Quasinormal modes in charged fluids at complex momentum},'' {\em JHEP} {\bf 10} (2020) 121, \href{http://www.arXiv.org/abs/2007.14418}{{\tt 2007.14418}}.

\bibitem{Grozdanov:2019uhi}
S.~Grozdanov, P.~K. Kovtun, A.~O. Starinets, and P.~Tadi\'c, ``{The complex life of hydrodynamic modes},'' {\em JHEP} {\bf 11} (2019) 097, \href{http://www.arXiv.org/abs/1904.12862}{{\tt 1904.12862}}.

\bibitem{Grozdanov:2021jfw}
S.~Grozdanov, A.~O. Starinets, and P.~Tadi\'c, ``{Hydrodynamic dispersion relations at finite coupling},'' {\em JHEP} {\bf 06} (2021) 180, \href{http://www.arXiv.org/abs/2104.11035}{{\tt 2104.11035}}.

\bibitem{Asadi:2021hds}
M.~Asadi, H.~Soltanpanahi, and F.~Taghinavaz, ``{Critical behaviour of hydrodynamic series},'' {\em JHEP} {\bf 05} (2021) 287, \href{http://www.arXiv.org/abs/2102.03584}{{\tt 2102.03584}}.

\bibitem{Taghinavaz:2023tog}
F.~Taghinavaz, ``{Relativistic hydrodynamics with phase transition},'' \href{http://www.arXiv.org/abs/2309.14773}{{\tt 2309.14773}}.

\bibitem{Grozdanov:2020koi}
S.~Grozdanov, ``{Bounds on transport from univalence and pole-skipping},'' {\em Phys. Rev. Lett.} {\bf 126} (2021), no.~5, 051601, \href{http://www.arXiv.org/abs/2008.00888}{{\tt 2008.00888}}.

\bibitem{Duren:2010pm}
P.~L. Duren, {\em {Univalent functions}}.
\newblock Grundlehren der mathematischen Wissenschaften. Springer New York, NY, 2011.

\bibitem{lehto2011univalent}
O.~Lehto, {\em Univalent Functions and Teichm{\"u}ller Spaces}.
\newblock Graduate Texts in Mathematics. Springer New York, 2011.

\bibitem{Baggioli:2022uqb}
M.~Baggioli, S.~Grieninger, S.~Grozdanov, and Z.~Lu, ``{Aspects of univalence in holographic axion models},'' {\em JHEP} {\bf 11} (2022) 032, \href{http://www.arXiv.org/abs/2205.06076}{{\tt 2205.06076}}.

\bibitem{Haldar:2021rri}
P.~Haldar, A.~Sinha, and A.~Zahed, ``{Quantum field theory and the Bieberbach conjecture},'' {\em SciPost Phys.} {\bf 11} (2021) 002, \href{http://www.arXiv.org/abs/2103.12108}{{\tt 2103.12108}}.

\bibitem{Grozdanov:2018fic}
S.~Grozdanov, A.~Lucas, and N.~Poovuttikul, ``{Holography and hydrodynamics with weakly broken symmetries},'' {\em Phys. Rev. D} {\bf 99} (2019), no.~8, 086012, \href{http://www.arXiv.org/abs/1810.10016}{{\tt 1810.10016}}.

\bibitem{Bemfica:2017wps}
F.~S. Bemfica, M.~M. Disconzi, and J.~Noronha, ``{Causality and existence of solutions of relativistic viscous fluid dynamics with gravity},'' {\em Phys. Rev. D} {\bf 98} (2018), no.~10, 104064, \href{http://www.arXiv.org/abs/1708.06255}{{\tt 1708.06255}}.

\bibitem{Kovtun:2019hdm}
P.~Kovtun, ``{First-order relativistic hydrodynamics is stable},'' {\em JHEP} {\bf 10} (2019) 034, \href{http://www.arXiv.org/abs/1907.08191}{{\tt 1907.08191}}.

\bibitem{Jensen:2012jh}
K.~Jensen, M.~Kaminski, P.~Kovtun, R.~Meyer, A.~Ritz, and A.~Yarom, ``{Towards hydrodynamics without an entropy current},'' {\em Phys. Rev. Lett.} {\bf 109} (2012) 101601, \href{http://www.arXiv.org/abs/1203.3556}{{\tt 1203.3556}}.

\bibitem{Hoult:2020eho}
R.~E. Hoult and P.~Kovtun, ``{Stable and causal relativistic Navier-Stokes equations},'' {\em JHEP} {\bf 06} (2020) 067, \href{http://www.arXiv.org/abs/2004.04102}{{\tt 2004.04102}}.

\bibitem{Taghinavaz:2020axp}
F.~Taghinavaz, ``{Causality and Stability Conditions of a Conformal Charged Fluid},'' {\em JHEP} {\bf 08} (2020) 119, \href{http://www.arXiv.org/abs/2004.01897}{{\tt 2004.01897}}.

\end{thebibliography}\endgroup
\end{document}